\documentclass[vecphys]{svmult}

\usepackage{makeidx}         
\usepackage{graphicx}        
\usepackage{multicol}        
\usepackage[bottom]{footmisc}
\newcommand{\anti}[1]{\overline{#1}}

\makeindex             

\begin{document}

\title*{The hyperon-nucleon interaction: conventional versus effective
field theory approach}
\titlerunning{Hyperon-nucleon interaction} 
\author{J. Haidenbauer\inst{1}\and
Ulf-G. Mei{\ss}ner\inst{1,2}\and
A. Nogga\inst{1}\and
H. Polinder\inst{1}
}
\institute{Institut f{\"u}r Kernphysik (Theorie), Forschungszentrum J{\"u}lich, 
D-52425 J{\"u}lich, Germany
\and 
Helmholtz-Institut f\"ur Strahlen- und Kernphysik (Theorie),
Universit\"at Bonn, Nu\ss allee 14-16, D-53115 Bonn, Germany
}
%
%
\maketitle

Hyperon-nucleon interactions are presented that are derived either in the 
conventional meson-exchange picture or within leading order chiral effective 
field theory.  The chiral potential consists of one-pseudoscalar-meson exchanges 
and non-derivative four-baryon contact terms. 
With regard to meson-exchange hyperon-nucleon models we focus on the new potential of 
the J\"ulich group, whose most salient feature is that the contributions 
in the scalar--isoscalar ($\sigma$) and vector--isovector ($\rho$) exchange channels 
are constrained by a microscopic model of correlated $\pi\pi$ and $K \anti{K}$ 
exchange.

\section{Introduction}
\label{chap:1}

For several decades the meson-exchange picture provided the only 
practicable and systematic approach to the description of hadronic
reactions in the low- and medium-energy regime. Specifically, for
the fundamental nucleon-nucleon ($NN$) interaction rather precise
quantitative results could be achieved with meson-exchange models
\cite{MHE,Mac01}.  Moreover, utilizing for example 
${\rm SU(3)}_f$ (flavor) symmetry or $G$-parity 
arguments, within the meson-exchange framework, interaction models for the 
hyperon-nucleon ($YN$) \cite{MRS89,Hol89,Reu94,YNN1,Rij99,Tom01,Hai05,Rij06}
or nucleon-antinucleon ($N\anti{N}$) \cite{Kle02} systems could 
be constructed consistently. 
However, over the last 10 years or so a new powerful tool has emerged,
namely chiral perturbation theory or, generally speaking, effective
field theory (EFT). 
The main advantage of this scheme is that there is an underlying
power counting that allows to improve calculations systematically by 
going to higher orders and, at the same time, provides theoretical
uncertainties. In addition, it is possible to derive two- and corresponding
three-body forces as well as external current operators in a consistent way.
For reviews we refer to \cite{Bed02,Kap05,Epelbaum:2005pn}.
 
Recently the $NN$ interaction has been described
to a high precision using chiral EFT \cite{Epe05} (see also \cite{Entem:2003ft}).       
In that work, the power counting is applied to the $NN$ potential, as
originally proposed by Weinberg \cite{Wei90,Wei91}. The $NN$ potential consists 
of pion exchanges and a series of contact interactions with an increasing number
of derivatives to parameterize the shorter ranged part of the $NN$ force. 
A regularized Lippmann-Schwinger
equation is solved to calculate observable quantities. Note that in contrast
to the original Weinberg scheme, the effective potential is made explicitly
energy-independent as it is important for applications in few-nucleon
systems (for details, see \cite{Epe98}).

Contrary to the $NN$ system, there are very few investigations of 
the $YN$ interaction using EFT. Hyperon and nucleon mass shifts in
nuclear matter, using chiral perturbation theory, have been studied in
\cite{Sav96}. These authors used a chiral interaction containing four-baryon contact
terms and pseudoscalar-meson exchanges. Recently, the hypertriton and $\Lambda
d$ scattering were investigated in the framework of an EFT with contact
interactions only \cite{Ham02}.
Korpa et al. \cite{Kor01} performed a next-to-leading order (NLO) EFT analysis of
$YN$ scattering and hyperon mass shifts in nuclear matter. Their tree-level
amplitude contains four-baryon contact terms; pseudoscalar-meson exchanges
were not considered explicitly, but ${\rm SU(3)}_f$ breaking by meson masses
was modeled by incorporating dimension two terms coming from one-pion
exchange. The full scattering amplitude was calculated using the
Kaplan-Savage-Wise resummation scheme \cite{Kap98}. The $YN$ scattering
data were described successfully for laboratory momenta below 200 MeV, using
12 free parameters. Some aspects of strong $\Lambda N$ scattering in EFT
and its relation to various formulations of lattice QCD are
discussed in \cite{Beane:2003yx}. Finally, in this context we note that first 
lattice QCD results on the $YN$ interaction have appeared \cite{Bea06}.

In this review we describe a a recent application of the scheme used in 
\cite{Epe05} to the $YN$ interaction by the Bonn-J\"ulich group \cite{Pol06}. 
Analogous to the $NN$ potential, at leading order (LO) in the power
counting, the $YN$ potential consists of pseudoscalar-meson (Goldstone boson)
exchanges and of four-baryon contact terms, where each of these two contributions
is constrained via ${\rm SU(3)}_f$ symmetry. 
The results achieved by us within this approach are confronted with the available 
$YN$ data and they are also compared with predictions of a new conventional 
meson-exchange $YN$ model, 
developed likewise by the J\"ulich group \cite{Hai05}, whose most salient feature is 
that the contributions in the scalar--isoscalar ($\sigma$) and vector--isovector 
($\rho$) exchange channels are constrained by a microscopic model of correlated 
$\pi\pi$ and $K \anti{K}$ exchange. Results of the Nijmegen $YN$ model NSC97f 
\cite{Rij99} are presented too. 

The contents of this review are as follows. In Section 2 we discuss some general properties of the
coupled $\Lambda N$ and $\Sigma N$ systems. We also introduce the 
coupled-channels Lippmann-Schwinger equation that is solved for obtaining the 
reaction amplitude. 
The effective potential in leading order chiral EFT is developed
in Section~\ref{chap:2}. Here we first give a brief recollection of the underlying power 
counting for the effective potential and then investigate the ${\rm SU(3)}_f$ structure of the 
four-baryon contact interactions. The lowest order ${\rm SU(3)}_f$-invariant contributions
from pseudoscalar meson exchange are derived too. 
Some general remarks about meson-exchange potentials of the $YN$ interaction are given in 
Section~\ref{chap:3}. We also provide a more specific description of the new meson-exchange 
potential of the J\"ulich group \cite{Hai05}, 
where we focus on the utilized model of correlated $\pi\pi$ and $K\anti{K}$ exchange. 
Results of both interactions for low-energy $YN$ cross sections are presented in Section \ref{chap:6}. 
We show the empirical and calculated total cross sections, differential cross sections and give the 
values for the scattering lengths. 
Also, predictions for some $YN$ phase shifts are presented and results for binding energies of light
hypernuclei are listed. 
The review closes with a summary and an outlook for future investigations. 

\section{The scattering equation}
\label{chap:7}
In the meson-meson and meson-baryon sector, chiral interactions can be
treated perturbatively in powers of a low-energy scale (chiral
perturbation theory). This is not the case for the
baryon-baryon sector, otherwise there could be no bound states, such as
the deuteron. Weinberg \cite{Wei91} realized that an additional scale
arises from intermediate states with only two nucleons, which requires a
modification of the power counting. He proposed to apply the techniques
of chiral perturbation theory to derive an effective potential, $V$, and not 
directly the scattering amplitude. This effective potential is defined as 
the sum of all irreducible diagrams. The effective potential $V$ is then put 
into a Lippmann-Schwinger equation to obtain the reaction or scattering amplitude,
\begin{eqnarray}
T = V + V G T \ ,
\label{eq:ls}
\end{eqnarray}
where $G$ is the non-relativistic free two-body Green's function. 
Solving the scattering equation (\ref{eq:ls}) also implies that the 
reaction amplitude $T$ fulfills two-body unitarity. 

Treating the Lippmann-Schwinger equation for the $YN$ system is 
more involved than for the $NN$ system. Since the mass difference
between the $\Lambda$ and $\Sigma$ hyperons is only about 75
MeV the possible coupling between the $\Lambda N$ and $\Sigma N$
systems needs to be taken into account. Moreover, for a sensible 
comparison of the results with experiments it is preferable to
solve the scattering equation in the particle basis because then 
the Coulomb interaction in the charged channels can be incorporated.
Here we use the method originally introduced by Vincent and 
Phatak \cite{Vin74} that was e.g. also applied in the EFT studies
of the $NN$ interaction \cite{WME01}. Furthermore, the particle basis
allows to implement the correct physical thresholds of the various 
$\Sigma N$ channels. 
To facilitate the latter aspect we also use relativistic kinematics 
for relating the total energy $\sqrt{s}$ to the c.m. momenta in the 
various $YN$ channels in the actual calculations, cf. \cite{Hai05}. 
Note that the interaction potentials themselves are calculated in 
the isospin basis. 

The concrete particle channels that couple for a specific charge
$Q$ are
\begin{eqnarray}
Q=+2: &&\Sigma^+p \nonumber\\
Q=+1: &&\Lambda p, \Sigma^+n, \Sigma^0 p \nonumber\\
Q=\phantom{+}0 : &&\Lambda n, \Sigma^0n, \Sigma^- p \nonumber\\
Q=-1: &&\Sigma^-n 
\end{eqnarray}
Therefore, e.g., for $Q=0$ the quantities in Eq. (\ref{eq:ls}) are then 
$3\times 3$ matrices,
\begin{eqnarray}
&&
V = 
\left(
\begin{array}{ccc}
V_{\Lambda n \to \Lambda n} &V_{\Lambda n \to \Sigma^0 n}  &V_{\Lambda n \to
\Sigma^- p}  \\
V_{\Sigma^0 n \to \Lambda n} &V_{\Sigma^0 n \to \Sigma^0 n}  &V_{\Sigma^0 n \to
\Sigma^- p}  \\
V_{\Sigma^- p \to \Lambda n} &V_{\Sigma^- p \to \Sigma^0 n}  &V_{\Sigma^- p \to
\Sigma^- p}
\end{array}
\right)
 \ ,  
\end{eqnarray}
and analogously for $T$ while the Green's function is a diagonal matrix, 
\begin{eqnarray}
&&
G = 
\left(
\begin{array}{ccc}
G_{\Lambda n} &0 &0  \\
0 &G_{\Sigma^0 n}  & 0 \\
0 & 0 &G_{\Sigma^- p}
\end{array}
\right)
 \ .  
\end{eqnarray}
Explicitly, $G_i$ is given by 
\begin{equation}
G_i = \left[ {p_i^2-{\bf p}'^2 \over {2\mu_i}} + i\varepsilon \right]^{-1} \ ,
\label{Green}
\end{equation}
where $\mu_i = M_{Y_i}M_{N_i}/(M_{Y_i}+M_{N_i})$ is the reduced mass and ${\bf p}'$ the
c.m. momentum in the intermediate $Y_iN_i$ channel. $p_i = p_i(\sqrt{s})$ denotes 
the modulus of the
on-shell momentum in the intermediate $Y_iN_i$ state defined by
$\sqrt{s} = \sqrt{M_{Y_i}^2 + p_i^2} + \sqrt{M_{N_i}^2 + p_i^2}$.

\section{Hyperon-nucleon potential based on effective field theory}
\label{chap:2}

In this Section, we construct in some detail the effective chiral
$YN$ potential at leading order in the (modified) Weinberg
power counting. This power counting is briefly recalled first. Then,
we construct the minimal set of non-derivative four-baryon interactions
and derive the formulae for the one-Goldstone-boson-exchange contributions.

\subsection{Power counting}
\label{chap:2.1}

In our work \cite{Pol06} we apply the power counting to the effective $YN$ 
potential $V$ which is then injected into a Lippmann-Schwinger 
equation (\ref{eq:ls}) to generate the bound and scattering states. 
The various terms in the effective potential are ordered according to
\begin{equation}
V \equiv V(Q,g,\mu) = \sum_\nu Q^\nu \, {\mathcal V}_\nu (Q/\mu ,g)~,
\end{equation}
where $Q$ is the soft scale (either a baryon three-momentum, a
Goldstone boson four-momentum  or a Goldstone boson mass), $g$ is a generic
symbol for the pertinent low--energy constants, $\mu$ a regularization
scale, ${\mathcal V}_\nu$ is a function of order one, and $\nu \ge 0$ is the chiral power.
It can be expressed as 
\cite{Epelbaum:2005pn} 
\begin{eqnarray}\label{eq:power}
\nu &=& 2 - B + 2L + \sum_i v_i \,\Delta_i ~,\nonumber\\
\Delta_i &=& d_i + {\displaystyle\frac{1}{2}}\, b_i  - 2~,
\end{eqnarray}
with $B$ the number of incoming (outgoing) baryon fields, $L$ counts the
number of Goldstone boson loops, and $v_i$ is the number of vertices with
dimension $\Delta_i$. The vertex dimension is expressed in terms of 
derivatives (or Goldstone boson masses) $d_i$  and the number of internal
baryon fields $b_i$ at the vertex under consideration. The LO
potential is given by $\nu = 0$, with $B=2$, $L=0$ and $\Delta_i = 0$. Using
Eq.~(\ref{eq:power}) it is easy to see that this condition is fulfilled
for two types of interactions -- a) non-derivative four-baryon contact terms 
with $b_i = 4$ and $d_i = 0$ and b) one-meson exchange diagrams with the
leading meson-baryon derivative vertices allowed by chiral symmetry ($b_i = 2,
d_i = 1$). At LO, the effective potential is entirely given by these two
types of contributions. 

\subsection{The four-baryon contact terms}
\label{chap:2.2}
Let us start with briefly recalling the situation for the $NN$ interactions. 
The LO contact term for the $NN$ interactions 
is given by e.g. \cite{Wei90,Epe98}
\begin{eqnarray}
{\mathcal L}&=&C_i\left(\bar{N}\Gamma_i N\right)\left(\bar{N}\Gamma_i N\right)\ ,
\label{eq:2.1}
\end{eqnarray}
where $\Gamma_i$ are the usual elements of the Clifford algebra \cite{Bjo65}
\begin{equation}
\Gamma_1=1 \, , \,\, 
\Gamma_2=\gamma^\mu \, , \,\,   
\Gamma_3=\sigma^{\mu\nu} \, , \,\,  
\Gamma_4=\gamma^\mu\gamma_5  \, , \,\, 
\Gamma_5=\gamma_5 \,\, ,
\label{eq:2.2}
\end{equation}
$N$ are the Dirac spinors of the nucleons and $C_i$ are the so-called low-energy constants (LECs). 
The small components of the nucleon spinors do not contribute to the LO contact interactions. 
Considering the large components only, the LO contact term, Eq. (\ref{eq:2.1}), becomes
\begin{eqnarray}
{\mathcal L}&=&
-\frac{1}{2}C_S\left(\varphi^\dagger_N\varphi_N\right)\left(\varphi^\dagger_N\varphi_N\right)-\frac{1}{2}C_T\left(\varphi^\dagger_N \mbox{\boldmath $\sigma$} \varphi_N\right)\left(\varphi^\dagger_N \mbox{\boldmath $\sigma$} \varphi_N\right)\ ,
\label{eq:2.3}
\end{eqnarray}
where $\varphi_N$ denotes the large component of the Dirac spinor and
$C_S$ and $C_T$ are the LECs that need to be determined by fitting to the experimental data. 

\begin{figure}[t]
\begin{center}
\resizebox{\textwidth}{!}{\includegraphics*[2cm,21.0cm][18cm,26cm]{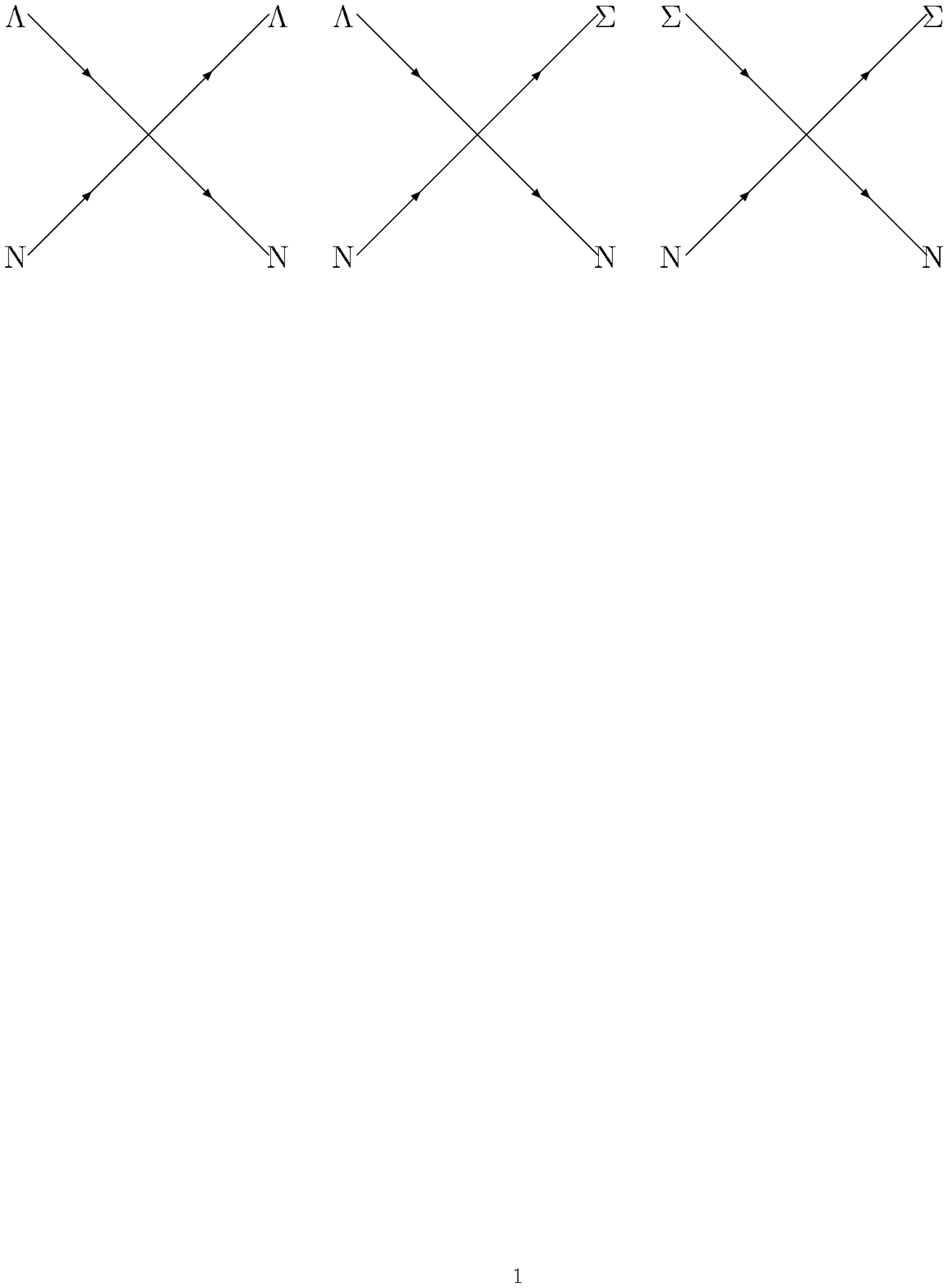}}
\end{center}
\caption{Lowest order contact terms for hyperon-nucleon interactions}
\label{fig:2.0}
\end{figure}

In the case of the $YN$ interaction we will consider a similar but ${\rm SU(3)}_f$ invariant 
coupling. 
The LO contact terms for the octet baryon-baryon interactions, that are Hermitian 
and invariant under Lorentz transformations, are given by the ${\rm SU(3)}_f$ invariants,
\begin{eqnarray}
{\mathcal L}^1 &=& C^1_i \left<\bar{B}_a\bar{B}_b\left(\Gamma_i B\right)_b\left(\Gamma_i B\right)_a\right>\ ,\quad 
{\mathcal L}^2=C^2_i \left<\bar{B}_a\left(\Gamma_i B\right)_a\bar{B}_b\left(\Gamma_i B\right)_b\right>\ ,\nonumber \\
{\mathcal L}^3&=&C^3_i \left<\bar{B}_a\left(\Gamma_i B\right)_a\right>\left<\bar{B}_b\left(\Gamma_i B\right)_b\right>, \
{\mathcal L}^4=C^4_i \left<\bar{B}_a\bar{B}_b\left(\Gamma_i
    B\right)_a\left(\Gamma_i B\right)_b\right>\ , \nonumber \\
{\mathcal L}^5 &=&C^5_i \left<\bar{B}_a\left(\Gamma_i B\right)_b\bar{B}_b\left(\Gamma_i B\right)_a\right>\ ,\quad
{\mathcal L}^6 = C^6_i \left<\bar{B}_a\left(\Gamma_i B\right)_b\right>\left<\bar{B}_b\left(\Gamma_i B\right)_a\right>\ ,\nonumber \\
{\mathcal L}^7&=&C^7_i \left<\bar{B}_a\left(\Gamma_i B\right)_a\left(\Gamma_i B\right)_b\bar{B}_b\right>\ ,\quad
{\mathcal L}^8 = C^8_i \left<\bar{B}_a\left(\Gamma_i B\right)_b\left(\Gamma_i B\right)_a\bar{B}_b\right>\ ,\nonumber \\
{\mathcal L}^9&=&C^9_i \left<\bar{B}_a\bar{B}_b\right>\vphantom{\bar{B}_a}\left<\left(\Gamma_i B\right)_a\left(\Gamma_i B\right)_b\vphantom{\bar{B}_a}\right>\ .
\label{eq:2.4}
\end{eqnarray}
Here $a$ and $b$ denote the Dirac indices of the particles, $B$ is the usual irreducible octet representation of ${\rm SU(3)}_f$ given by
\begin{eqnarray}
B&=&
\left(
\begin{array}{ccc}
\frac{\Sigma^0}{\sqrt{2}}+\frac{\Lambda}{\sqrt{6}} & \Sigma^+ & p \\
\Sigma^- & \frac{-\Sigma^0}{\sqrt{2}}+\frac{\Lambda}{\sqrt{6}} & n \\
-\Xi^- & \Xi^0 & -\frac{2\Lambda}{\sqrt{6}}
\end{array}
\right) \ ,
\label{eq:2.5}
\end{eqnarray}
and the brackets in (\ref{eq:2.4}) denote taking the trace in the three-dimensional flavor space. The Clifford algebra elements are here 
actually diagonal $3\times 3$ matrices in flavor space. Term 9 in Eq. (\ref{eq:2.4}) can be eliminated using a Cayley-Hamilton identity
\begin{eqnarray}
&&-\left<\bar{B}_a\bar{B}_b\left(\Gamma_i B\right)_a\left(\Gamma_i B\right)_b\right>+\left<\bar{B}_a\bar{B}_b\left(\Gamma_i B\right)_b\left(\Gamma_i B\right)_a\right> \nonumber \\
&&-\frac{1}{2}\left<\bar{B}_a\left(\Gamma_i B\right)_b\bar{B}_b\left(\Gamma_i B\right)_a\right>+\frac{1}{2}\left<\bar{B}_a\left(\Gamma_i B\right)_a\bar{B}_b\left(\Gamma_i B\right)_b\right> \nonumber \\
&&=\frac{1}{2}\left<\bar{B}_a\left(\Gamma_i B\right)_a\right>\left<\bar{B}_b\left(\Gamma_i B\right)_b\right>-\frac{1}{2}\left<\bar{B}_a\left(\Gamma_i B\right)_b\right>\left<\bar{B}_b\left(\Gamma_i B\right)_a\right> \nonumber \\
&&-\frac{1}{2}\left<\bar{B}_a\bar{B}_b\right>\vphantom{\bar{B}_a}\left<\left(\Gamma_i B\right)_a\left(\Gamma_i B\right)_b\vphantom{\bar{B}_a}\right>\ .
\end{eqnarray}
Making use of the trace property $\left<AB\right>=\left<BA\right>$, we see that the terms 7 and 8 in Eq. (\ref{eq:2.4}) are equivalent to the terms 1 and 4 respectively. Also making use of the Fierz theorem, see e.g. \cite{Che84}, one can show that the terms 4, 5 and 6 are equivalent to the terms 1, 2 and 3, respectively. 
So, the minimal set of non-derivative four baryon contact interactions is given by ${\mathcal L}^1$, ${\mathcal L}^2$ and ${\mathcal L}^3$. Writing these interaction Lagrangians explicitly in the isospin basis we find for the $NN$ and $YN$ interactions
\begin{eqnarray}
{\mathcal L}^1&=&C^1_i\left\{ \frac{1}{6}\left[ 5\left(\bar{\Lambda}\Gamma_i \Lambda\right)\left(\bar{N}\Gamma_iN\right)-4\left(\bar{N}\Gamma_i\Lambda\right)\left(\bar{\Lambda}\Gamma_i N\right) \right]  \right. \nonumber \\
&&+\frac{1}{2}\left[ \left(\bar{\mbox{\boldmath $\Sigma$}}\cdot\Gamma_i\mbox{\boldmath $\Sigma$}\right)\left(\bar{N}\Gamma_iN\right) + i\left(\bar{\mbox{\boldmath $\Sigma$}}\times\Gamma_i\mbox{\boldmath $\Sigma$}\right)\cdot\left(\bar{N}\mbox{\boldmath $\tau$}\Gamma_iN\right)  \right] \nonumber \\
&&+ \frac{1}{\sqrt{12}}\left[ \left\{\left(\bar{N}\mbox{\boldmath $\tau$}\Gamma_iN\right)\cdot\left(\bar{\Lambda}\Gamma_i\mbox{\boldmath $\Sigma$}\right)+H.c.\right\} \right. \nonumber \\
&&\left.\left. -2\left\{\left(\bar{N}\Gamma_i\mbox{\boldmath $\Sigma$}\right)\cdot\left(\bar{\Lambda}\mbox{\boldmath $\tau$}\Gamma_iN\right)+H.c.\right\} \right] \vphantom{\frac{1}{6}}\right\} 
\ , \nonumber \\
{\mathcal L}^2&=&C^2_i\left\{\vphantom{\frac{1}{\sqrt{3}}} \frac{1}{3}\left[ 4\left(\bar{\Lambda}\Gamma_i \Lambda\right)\left(\bar{N}\Gamma_iN\right)+\left(\bar{N}\Gamma_i\Lambda\right)\left(\bar{\Lambda}\Gamma_i N\right) \right]  \right. \nonumber \\
&&+\left[ \left(\bar{N}\Gamma_i\mbox{\boldmath $\Sigma$}\right)\cdot\left(\bar{\mbox{\boldmath $\Sigma$}}\Gamma_i N\right) + i\left(\bar{N}\Gamma_i \mbox{\boldmath $\Sigma$}\right)\cdot\left(\bar{\mbox{\boldmath $\Sigma$}}\times\mbox{\boldmath $\tau$}\Gamma_i N\right)  \right] \nonumber \\
&&\left. +\frac{1}{\sqrt{3}}\left[ \left(\bar{N}\Gamma_i \mbox{\boldmath $\Sigma$}\right)\cdot\left(\bar{\Lambda}\mbox{\boldmath $\tau$}\Gamma_i N\right)+H.c. \right] +\left(\bar{N}\Gamma_iN\right)\left(\bar{N}\Gamma_iN\right)\right\} 
\ , \nonumber \\
{\mathcal L}^3&=&C^3_i\left\{
  2\left(\bar{\Lambda}\Gamma_i\Lambda\right)\left(\bar{N}\Gamma_i
    N\right)+2\left(\bar{\mbox{\boldmath
        $\Sigma$}}\cdot\Gamma_i\mbox{\boldmath
      $\Sigma$}\right)\left(\bar{N}\Gamma_i N\right) 
+\left(\bar{N}\Gamma_iN\right)\left(\bar{N}\Gamma_iN\right)  \right\}\ . \nonumber\\&&
\label{eq:2.4a}
\end{eqnarray}
Here $H.c.$ denotes the Hermitian conjugate of the specific term. Also $\Lambda$ is an isoscalar, 
$N$ and $\Xi$ are isospinors and $\mbox{\boldmath$\Sigma$}$ is an isovector:
\begin{equation}
N=\left(\begin{array}{r}p\\n\end{array}\right)\ ,\ \ \Xi=\left(\begin{array}{l}\Xi^0\\\Xi^-\end{array}\right)\ ,\ \ 
\mbox{\boldmath $\Sigma$}=\left(\begin{array}{l}\Sigma^+\\\Sigma^0\\\Sigma^-\end{array}\right)\ .
\label{eq:2.4b}
\end{equation}
The LO $YN$ contact terms given by these 
Lagrangians are shown diagrammatically in Fig. \ref{fig:2.0}.
Considering again only the large components of the Dirac spinors, similar to Eq. (\ref{eq:2.3}), we 
arrive at six contact constants ($C^1_S$, $C^1_T$, $C^2_S$, $C^2_T$, $C^3_S$ and $C^3_T$) for 
the interactions in the various $BB\to BB$ channels. The LO contact potentials resulting from 
the above Lagrangians have the form
\begin{eqnarray}
V^{BB\to BB}&=&C_S^{BB\to BB}+C_T^{BB\to BB}\ \mbox{\boldmath$\sigma_1$}\cdot\mbox{\boldmath$\sigma_2$} \ .
\label{eq11}
\end{eqnarray}
Projecting the LO contact potential on the partial waves, 
for details see, e.g., Ref. \cite{Rij95}, one finds the following contributions. The $NN$ partial wave potentials are
\begin{eqnarray}
V^{NN}_{1S0}&=&4\pi\left[2\left(C^2_S-3C^2_T\right)+2\left(C^3_S-3C^3_T\right)\right]=V^{27} , \nonumber \\
V^{NN}_{3S1}&=&4\pi\left[2\left(C^2_S+C^2_T\right)+2\left(C^3_S+C^3_T\right)\right]=V^{10^*} .
\label{eq:2.13a}
\end{eqnarray}
The partial wave potentials for $\Lambda N \rightarrow \Lambda N$ are 
\begin{eqnarray}
V^{\Lambda\Lambda}_{1S0}&=&4\pi\left[\frac{1}{6}\left(C^1_S-3C^1_T\right)+\frac{5}{3}\left(C^2_S-3C^2_T\right)+2\left(C^3_S-3C^3_T\right)\right] 
\nonumber \\
&=&\frac{1}{10}\left(9V^{27}+V^{8s}\right) , \nonumber \\
V^{\Lambda\Lambda}_{3S1}&=&4\pi\left[\frac{3}{2}\left(C^1_S+C^1_T\right)+\left(C^2_S+C^2_T\right)+2\left(C^3_S+C^3_T\right)\right]
 \nonumber \\
&=&\frac{1}{2}\left(V^{8a}+V^{10^*}\right) ,
\label{eq:2.13}
\end{eqnarray}
where here and in the following we introduced the shorthand notation ``$\Lambda \Lambda$'' instead of 
$\Lambda N\to \Lambda N$, etc., for labelling the interaction potentials and the corresponding contact terms. 
For isospin-3/2 $\Sigma N\rightarrow \Sigma N$ one gets 
\begin{eqnarray}
V^{\Sigma\Sigma}_{1S0}&=&4\pi\left[2\left(C^2_S-3C^2_T\right)+2\left(C^3_S-3C^3_T\right)\right]=V^{27} , \nonumber \\
V^{\Sigma\Sigma}_{3S1}&=&4\pi\left[-2\left(C^2_S+C^2_T\right)+2\left(C^3_S+C^3_T\right)\right]=V^{10} ,
\label{eq:2.14}
\end{eqnarray}
for isospin-1/2 $\Sigma N\rightarrow \Sigma N$
\begin{eqnarray}
\widetilde{V}^{\Sigma\Sigma}_{1S0}&=&4\pi\left[\frac{3}{2}\left(C^1_S-3C^1_T\right)-\left(C^2_S-3C^2_T\right)+2\left(C^3_S-3C^3_T\right)\right] \nonumber \\
&=&\frac{1}{10}\left(V^{27}+9V^{8s}\right) , \nonumber \\
\widetilde{V}^{\Sigma\Sigma}_{3S1}&=&4\pi\left[\frac{3}{2}\left(C^1_S+C^1_T\right)+\left(C^2_S+C^2_T\right)+2\left(C^3_S+C^3_T\right)\right] \nonumber \\
&=&\frac{1}{2}\left(V^{8a}+V^{10^*}\right) ,
\label{eq:2.15}
\end{eqnarray}
and for $\Lambda N\rightarrow \Sigma N$
\begin{eqnarray}
V^{\Lambda\Sigma}_{1S0}&=&4\pi\left[\frac{1}{2}\left(C^1_S-3C^1_T\right)-\left(C^2_S-3C^2_T\right)\right]=\frac{3}{10}\left(-V^{27}+V^{8s}\right) , \nonumber \\
V^{\Lambda\Sigma}_{3S1}&=&4\pi\left[-\frac{3}{2}\left(C^1_S+C^1_T\right)+\left(C^2_S+C^2_T\right)\right]=\frac{1}{2}\left(-V^{8a}+V^{10^*}\right) .
\label{eq:2.16}
\end{eqnarray}
The last relations in the previous Eqs. (\ref{eq:2.13a}) - (\ref{eq:2.16}) give explicitly 
the ${\rm SU(3)}_f$ representation of the potentials, see \cite{Swa63,Dov90}. We note that only 
5 of the $\{8\}\times\{8\}=\{27\}+\{10\}+\{10^*\}+\{8\}_s+\{8\}_a+\{1\}$ irreducible 
representations are relevant for $NN$ and $YN$ interactions, since the $\{1\}$ occurs 
only in the isospin zero $\Lambda\Lambda$, $\Xi N$ and $\Sigma\Sigma$ channels.  
Equivalently, the six contact terms, $C_S^1$, $C_T^1$, $C_S^2$, $C_T^2$, $C_S^3$, $C_T^3$, 
enter the $NN$ and $YN$ potentials in only 5 different combinations. 
These 5 contact terms need to be determined by a fit to the experimental data. 
Since the $NN$ data can not be described well with a LO EFT, see \cite{Wei90,Epe00a}, we 
will not consider the $NN$ interaction explicitly. 
Therefore, we are left with the $YN$ partial wave potentials
\begin{eqnarray}
&&
\begin{array}{lcllcl}
V^{\Lambda\Lambda}_{1S0}&=&C^{\Lambda\Lambda}_{1S0},& V^{\Lambda\Lambda}_{3S1}&=& C^{\Lambda\Lambda}_{3S1},\\
&&&&&\\
V^{\Sigma\Sigma}_{1S0}&=&C^{\Sigma\Sigma}_{1S0},& V^{\Sigma\Sigma}_{3S1}&=&C^{\Sigma\Sigma}_{3S1},\\
&&&&&\\
\widetilde{V}^{\Sigma\Sigma}_{1S0}&=&9C^{\Lambda\Lambda}_{1S0}-8C^{\Sigma\Sigma}_{1S0},& \widetilde{V}^{\Sigma\Sigma}_{3S1}&=&C^{\Lambda\Lambda}_{3S1},\\
&&&&&\\
V^{\Lambda\Sigma}_{1S0}&=&3\left(C^{\Lambda\Lambda}_{1S0}-C^{\Sigma\Sigma}_{1S0}\right),& V^{\Lambda\Sigma}_{3S1}&=&C^{\Lambda\Sigma}_{3S1}.
\end{array}
\label{eq17}
\end{eqnarray}
We have chosen to search for $C^{\Lambda \Lambda}_{1S0}$, $C^{\Lambda \Lambda}_{3S1}$, 
$C^{\Sigma \Sigma}_{1S0}$, $C^{\Sigma \Sigma}_{3S1}$, and $C^{\Lambda \Sigma}_{3S1}$ in the fitting procedure. The other partial wave potentials are then fixed by ${\rm SU(3)}_f$-symmetry.

\subsection{One pseudoscalar-meson exchange}
\label{chap:2.3}
The lowest order ${\rm SU(3)}_f$-invariant pseudoscalar-meson--baryon interaction Lagrangian 
is given by (see, e.g., \cite{Mei93}),
\begin{eqnarray}
{\mathcal L}&=&\left< i\bar{B}\gamma^\mu D_\mu B -M_0\bar{B}B+\frac{D}{2}\bar{B}\gamma^\mu\gamma_5 \left\{u_\mu,B\right\} +\frac{F}{2}\bar{B}\gamma^\mu\gamma_5 \left[u_\mu,B\right] \right> \ ,
\label{eq:3.1}
\end{eqnarray}
with $M_0$ the octet baryon mass in the chiral limit. 
There are two possibilities for coupling the axial vector $u_\mu$ to the baryon bilinear. The conventional coupling constants $F$ and $D$, used here, satisfy the relation $F+D=g_A\simeq 1.26$. The axial-vector strength $g_A$ is measured in neutron $\beta$--decay. 
The covariant derivative acting on the baryons is
\begin{eqnarray}
D_\mu B&=&\partial_\mu B+\left[\Gamma_\mu,B\right] \ ,\nonumber \\
\Gamma_\mu&=&\frac{1}{2}\left[u^\dagger\partial_\mu u+u\partial_\mu u^\dagger\right] \ , \nonumber \\
u^2&=&U=\exp (2iP/\sqrt{2}F_\pi) \ , \nonumber \\
u_\mu&=&i u^\dagger\partial_\mu U u^\dagger \ ,  
\label{eq:3.2}
\end{eqnarray}
where $F_\pi$ is the weak pion decay constant, $F_\pi =  92.4$ MeV, and $P$ is
the irreducible octet representation of ${\rm SU(3)}_f$ for the pseudoscalar
mesons (the Goldstone bosons)
\begin{eqnarray}
P&=&
\left(
\begin{array}{ccc}
\frac{\pi^0}{\sqrt{2}}+\frac{\eta}{\sqrt{6}} & \pi^+ & K^+ \\
\pi^- & \frac{-\pi^0}{\sqrt{2}}+\frac{\eta}{\sqrt{6}} & K^0 \\
K^- & \bar{K}^0 & -\frac{2\eta}{\sqrt{6}}
\end{array}
\right) \ .
\label{eq:3.3}
\end{eqnarray}
Symmetry breaking in the decay constants, e.g. $F_\pi \neq F_K$, formally
appears at NLO and will not be considered in the following. Writing the interaction Lagrangian explicitly in the isospin basis, we find
\begin{eqnarray}
{\mathcal L}&=&-f_{NN\pi}\bar{N}\gamma^\mu\gamma_5\mbox{\boldmath $\tau$}N\cdot\partial_\mu\mbox{\boldmath $\pi$} 
\nonumber \\
&&+if_{\Sigma\Sigma\pi}\bar{\mbox{\boldmath $ \Sigma$}}\gamma^\mu\gamma_5\times{\mbox{\boldmath $ \Sigma$}}\cdot\partial_\mu\mbox{\boldmath $\pi$} \nonumber \\
&&-f_{\Lambda\Sigma\pi}\left[\bar{\Lambda}\gamma^\mu\gamma_5{\mbox{\boldmath $ \Sigma$}}+\bar{\mbox{\boldmath $\Sigma$}}\gamma^\mu\gamma_5\Lambda\right]\cdot\partial_\mu\mbox{\boldmath $\pi$}
\nonumber \\
&&-f_{\Xi\Xi\pi}\bar{\Xi}\gamma^\mu\gamma_5\mbox{\boldmath $\tau$}\Xi\cdot\partial_\mu\mbox{\boldmath $\pi$} \nonumber \\
&&-f_{\Lambda NK}\left[\bar{N}\gamma^\mu\gamma_5\Lambda\partial_\mu K+\bar{\Lambda}\gamma^\mu\gamma_5N\partial_\mu K^\dagger\right]
\nonumber \\&&
-f_{\Xi\Lambda K}\left[\bar{\Xi}\gamma^\mu\gamma_5\Lambda\partial_\mu K_c+\bar{\Lambda}\gamma^\mu\gamma_5\Xi\partial_\mu K_c^\dagger\right]
\nonumber \\&&
-f_{\Sigma NK}\left[\bar{\mbox{\boldmath $ \Sigma$}}\cdot\gamma^\mu\gamma_5\partial_\mu K^\dagger\mbox{\boldmath $\tau$}N+\bar{N}\gamma^\mu\gamma_5\mbox{\boldmath $\tau$}\partial_\mu K\cdot{\mbox{\boldmath $ \Sigma$}}\right]
\nonumber \\&&
-f_{\Sigma \Xi K}\left[\bar{\mbox{\boldmath $ \Sigma$}}\cdot\gamma^\mu\gamma_5\partial_\mu K_c^\dagger\mbox{\boldmath $\tau$}\Xi+\bar{\Xi}\gamma^\mu\gamma_5\mbox{\boldmath $\tau$}\partial_\mu K_c\cdot{\mbox{\boldmath $ \Sigma$}}\right]
\nonumber \\&&
-f_{NN\eta_8}\bar{N}\gamma^\mu\gamma_5N\partial_\mu\eta
\nonumber \\&&
-f_{\Lambda\Lambda\eta_8}\bar{\Lambda}\gamma^\mu\gamma_5\Lambda\partial_\mu\eta
\nonumber \\&&
-f_{\Sigma\Sigma\eta_8}\bar{\mbox{\boldmath $ \Sigma$}}\cdot\gamma^\mu\gamma_5{\mbox{\boldmath $ \Sigma$}}\partial_\mu\eta
\nonumber \\&&
-f_{\Xi\Xi\eta_8}\bar{\Xi}\gamma^\mu\gamma_5\Xi\partial_\mu\eta \ .
\label{eq:3.7}
\end{eqnarray}
Here $\eta$ is an isoscalar, $K$ and $K_c$ are isospin doublets
\begin{equation}
K=\left(\begin{array}{r}K^+\\K^0\end{array}\right)\ ,\ \ K_c=\left(\begin{array}{r}\bar{K}^0\\-K^-\end{array}\right)\ ,
\label{eq:3.8}
\end{equation}
and $\mbox{\boldmath $ \pi$}$ is an isovector. The phases of the isovectors $\mbox{\boldmath $ \Sigma$}$ and 
$\mbox{\boldmath $ \pi$}$ are chosen such that \cite{Swa63} 
\begin{eqnarray}
\mbox{\boldmath $ \Sigma$}\cdot\mbox{\boldmath $ \pi$}&=&\Sigma^+\pi^-+\Sigma^0\pi^0+\Sigma^-\pi^+ \ .
\end{eqnarray}

\begin{figure}[t]
\begin{center}
\resizebox{\textwidth}{!}{\includegraphics*[2cm,16.5cm][13cm,26cm]{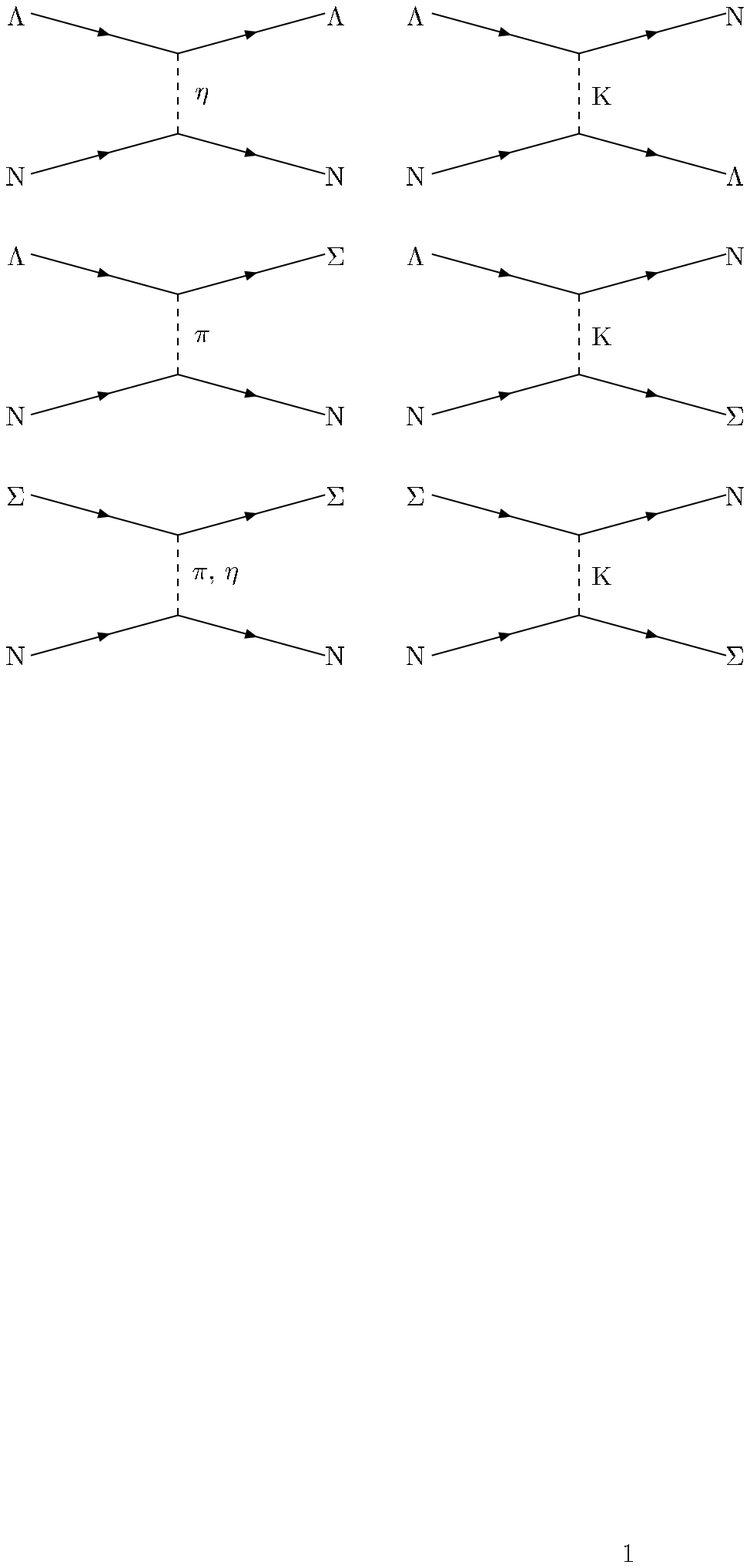}}
\end{center}
\caption{One-pseudoscalar-meson-exchange diagrams for hyperon-nucleon interactions.}
\label{fig:3.0}
\end{figure}

The interaction Lagrangian in Eq. (\ref{eq:3.7}) is invariant under $SU_f(3)$ transformations if the various coupling constants are expressed in terms of the coupling constant $f\equiv g_A/2F_\pi$ and the $F/(F+D)$-ratio $\alpha$ as \cite{Swa63},
\begin{equation}
\begin{array}{rlrlrl}
f_{NN\pi}  = & f, & f_{NN\eta_8}  = & \frac{1}{\sqrt{3}}(4\alpha -1)f, & f_{\Lambda NK} = & -\frac{1}{\sqrt{3}}(1+2\alpha)f, \\
f_{\Xi\Xi\pi}  = & -(1-2\alpha)f, &  f_{\Xi\Xi\eta_8}  = & -\frac{1}{\sqrt{3}}(1+2\alpha )f, & f_{\Xi\Lambda K} = & \frac{1}{\sqrt{3}}(4\alpha-1)f, \\
f_{\Lambda\Sigma\pi}  = & \frac{2}{\sqrt{3}}(1-\alpha)f, & f_{\Sigma\Sigma\eta_8}  = & \frac{2}{\sqrt{3}}(1-\alpha )f, & f_{\Sigma NK} = & (1-2\alpha)f, \\
f_{\Sigma\Sigma\pi}  = & 2\alpha f, &  f_{\Lambda\Lambda\eta_8}  = & -\frac{2}{\sqrt{3}}(1-\alpha )f, & f_{\Xi\Sigma K} = & -f.
\end{array}
\label{eq:3.9}
\end{equation}
The spin-space part of the one-pseudoscalar-meson-exchange potential resulting from the interaction Lagrangian 
Eq. (\ref{eq:3.7}) is in leading order, similar to the static one-pion-exchange potential 
(recoil and relativistic corrections give higher order contributions) in \cite{Epe98},
\begin{eqnarray}
V^{B_1B_2\to B_1'B'_2}&=&-f_{B_1B'_1P}f_{B_2B'_2P}\frac{\left(\mbox{\boldmath $\sigma$}_1\cdot{\bf k}\right)\left(\mbox{\boldmath $\sigma$}_2\cdot{\bf k}\right)}{{\bf k}^2+m^2_P}\ ,
\label{eq:3.10}
\end{eqnarray}
where $f_{B_1B'_1P}$, $f_{B_2B'_2P}$ are the appropriate coupling constants as given in Eq. (\ref{eq:3.9}) 
and $m_P$ is the actual mass of the exchanged pseudoscalar meson. Thus, the explicit SU(3) breaking 
reflected in the mass splitting between the pseudoscalar mesons is taken into account. 
With regard to the $\eta$ meson we identified its coupling with the octet value, i.e. the one for $\eta_8$, 
in our investigation \cite{Pol06}. (We will come back to that issue below.) 
We defined the transferred and average momentum, ${\bf k}$ and ${\bf q}$, in terms of the final and 
initial center-of-mass (c.m.) momenta of the baryons, ${\bf p}'$ and ${\bf p}$, as
\begin{equation}
{\bf k}={\bf p}'-{\bf p}\ ,\ \ \ {\bf q}=\frac{{\bf p}'+{\bf p}}{2} .
\label{eq:3.11}
\end{equation}
To find the complete LO one-pseudoscalar-meson-exchange potential one needs to multiply the 
potential in Eq.~(\ref{eq:3.10}) with the isospin factors given in Table~\ref{tab:3.1}.
\begin{table}[t]
\caption{The isospin factors for the various one--pseudoscalar-meson exchanges.}
\label{tab:3.1}
\centering
\begin{tabular}{|r|r|r|r|r|}
\hline
Channel &Isospin &$\pi$ &$K$ &$\eta$\\
\hline
$NN\rightarrow NN$ &$0$ &$-3$ &$0$ &$1$ \\
                   &$1$ &$1$  &$0$ &$1$ \\
\hline
$\Lambda N\rightarrow \Lambda N$ &$\frac{1}{2}$ &$0$ &$1$ &$1$ \\
\hline
$\Lambda N\rightarrow \Sigma N$ &$\frac{1}{2}$ &$-\sqrt{3}$ &$-\sqrt{3}$ &$0$ \\
\hline
$\Sigma N\rightarrow \Sigma N$ &$\frac{1}{2}$ &$-2$ &$-1$ &$1$ \\
                               &$\frac{3}{2}$ &$1$ &$2$ &$1$ \\
\hline
\end{tabular}
\end{table}
Fig. \ref{fig:3.0} shows the one-pseudoscalar-meson-exchange diagrams. 
Note that there is no contribution from one-pion exchange to the $\Lambda N
\to \Lambda N$ potential due to isospin conservation. Indeed, the longest
ranged contribution to this interaction is provided by (iterated) 
two-pion exchange via the process $\Lambda N \to \Sigma N \to \Lambda N$, 
generated by solving the Lippmann-Schwinger equation~(\ref{eq:ls}).  

\subsection{Determination of the low-energy constants}

The chiral EFT potential in the Lippmann-Schwinger equation is cut off with the regulator function 
\begin{equation}
f^\Lambda(p',p)=e^{-\left(p'^4+p^4\right)/\Lambda^4}\ ,
\label{regulator}
\end{equation}
in order to remove high-energy components of the baryon and pseudoscalar meson fields. For the cut-off 
$\Lambda$ we consider values between 550 and 700 MeV, this range is similar to the range used for chiral EFT $NN$ 
interactions \cite{Epe05,Epe02,Epe04}. The range is limited from below by the mass of the pseudoscalar 
mesons; since we do a LO calculation we do not expect a large plateau (i.e. a practically stable $\chi^2$ 
for varying $\Lambda$). 
  
For the fitting procedure we considered the empirical low-energy total cross sections 
published in Refs.~\cite{Sec68,Ale68,Eis71,Eng66}
and the inelastic capture ratio at rest \cite{Swa62}, in total 35 $YN$ data \cite{Pol06}. 
These data are also commonly used for determining the parameters of meson-exchange models.
The higher energy total cross sections and differential cross sections are then predictions of the LO chiral EFT, 
which contains five free parameters. The fits were done for fixed values of the cut--off mass ($\Lambda$) and of $\alpha$, 
the pseudoscalar $F/(F+D)$ ratio. For the latter we used the $SU(6)$ value: $\alpha=0.4$.
The five LECs $C^{\Lambda \Lambda}_{1S0}$, $C^{\Lambda \Lambda}_{3S1}$, $C^{\Sigma \Sigma}_{1S0}$, 
$C^{\Sigma \Sigma}_{3S1}$, and $ C^{\Lambda \Sigma}_{3S1}$ in Eq. (\ref{eq17}) were varied during 
the parameter search in order to fix the corresponding potentials.
The interaction in the other $YN$ partial waves (channels) are then determined by ${\rm SU(3)}_f$ symmetry. 
The values of the contact terms obtained in 
the fitting procedure for cut--off values between $550$ and $700$ MeV, are listed in Table \ref{tab:6.1}. 
\begin{table}[t]
\caption{The $YN$ $S$-wave contact terms for various cut--offs. The values of 
the LECs are in $10^4$ ${\rm GeV}^{-2}$; the values of $\Lambda$ in MeV. 
$\chi^2$ is the total chi squared for 35 $YN$ data.}
\label{tab:6.1}
\vspace{0.2cm}
\centering
\begin{tabular}{|r|rrrr|}
\hline
$\Lambda$& $550$& $600$& $650$& $700$  \\
\hline
$C^{\Lambda \Lambda}_{1S0}$ &$-.0466$ &$-.0403$ &$-.0322$ &$-.0304$\\
$C^{\Lambda \Lambda}_{3S1}$ &$-.0222$ &$-.0163$ &$-.0097$ &$-.0022$\\
$C^{\Sigma \Sigma}_{1S0}$   &$-.0766$ &$-.0763$ &$-.0757$ &$-.0744$\\
$C^{\Sigma \Sigma}_{3S1}$    &$.2336$  &$.2391$  &$.2392$  &$.2501$\\
$C^{\Lambda \Sigma}_{3S1}$  &$-.0016$ &$-.0019$  &$.0000$  &$.0035$\\
\hline
\hline
$\chi^2$& 29.6& 28.3& 30.3& 34.6\\
\hline
\end{tabular}
\end{table}

The fits were first done for the cut-off mass $\Lambda =600$ MeV. 
We remark that the $\Lambda N$ $S$-wave scattering lengths resulting for that cut-off were then 
kept fixed in the subsequent fits for the other cut--off values. 
We did this because the $\Lambda N$ scattering lengths are not well determined by the 
scattering data. As a matter of fact, not even the relative magnitude of the $\Lambda N$ triplet 
and singlet interaction can be constrained from the $YN$ data, 
but their strengths play an important role for the hypertriton binding energy \cite{YNN1}. 
Contrary to the $NN$ case, see, e.g. \cite{Epe00a}, the contact terms are in general not determined 
by a specific phase shift, because of the coupled particle channels in the $YN$ interaction. 
Furthermore, due to the limited accuracy and incompleteness of the $YN$ scattering data there is
no partial wave analysis. Therefore we have fitted the chiral EFT directly to the cross sections.
It is reassuring to see that the contact terms found in the parameter search are of similar 
magnitude as those obtained in the application of chiral EFT to the $NN$ interaction and,
specifically, they are of natural size \cite{Epelbaum:2005pn}. 

Note that we actually studied the dependence of our results on the pseudoscalar 
$F/(F+D)$ ratio $\alpha$ by
varying it within a range of 10 percent; after refitting the contact terms we
basically found an equally good description of the empirical data. 
An uncertainty in our calculation is the value of the $\eta$ coupling, since
we identified the physical $\eta$ with the octet $\eta$ as mentioned above. 
Therefore, we varied the $\eta$ 
coupling between zero and its octet value, but we found very little
influence on the description of the data (in fact, inclusion of the $\eta$
leads to a better plateau of the $\chi^2$ in the cut-off range considered).

\section{Hyperon-nucleon models based on the conventional meson-exchange 
picture}
\label{chap:3}

In the construction of conventional meson-exchange models of the $YN$
interaction usually one likewise assumes ${\rm SU(3)}_f$ symmetry for the
occurring coupling constants, and in some cases even the SU(6)
symmetry of the quark model \cite{Hol89,Reu94}. Indeed, in the derivation of the 
meson-exchange contributions one follows essentially the same procedure 
as outlined in Sect. \ref{chap:2.3} for the case of pseudoscalar mesons 
and, therefore, we do not present it here explicitly. Details can be 
found in Refs. \cite{Hol89,Rij95,NRS77}, for example. Of course, 
since besides the lowest pseudoscalar-meson multiplet also the 
exchanges of vector mesons ($\rho$, $\omega$, $K^*$), of scalar mesons 
($\sigma$, ...), or even of axial-vector mesons ($a_1(1270)$, ...) \cite{Rij06}
are included, one should keep in mind that the spin-space structure of
the corresponding Lagrangians that enter into Eq. (\ref{eq:3.1}) differ
and, accordingly, the final expressions for the corresponding 
contributions to the $YN$ interaction potentials differ too. 
Also we want to emphasize that even for pseudoscalar mesons
the final result for the interaction potentials differs, in general, 
from the expression given in Eq.~(\ref{eq:3.10}). Contrary to the 
chiral EFT approach, recoil and relativistic corrections are often 
kept in meson-exchange models because no power counting rules are applied.

The major conceptual difference between the various meson-exchange 
models consists in the treatment of the scalar-meson sector.  
This simply reflects the fact that, unlike for pseudoscalar 
and vector mesons, 
so far there is no general agreement about who are the actual members 
of the lowest lying scalar-meson SU(3) multiplet. (For a thorough
discussion on that issue and an overview of the extensive literature
we refer the reader to \cite{Kle04,Kal05} and references therein.)
Therefore, besides the question of the masses of the exchange particles 
it also remains unclear whether and how the relations for the coupling 
constants given in Eq. (\ref{eq:3.9}) should be applied.
As a consequence, different prescriptions for describing the scalar sector, 
whose contributions play a crucial role in any baryon-baryon interaction at 
intermediate ranges,  
were adopted by the various authors who published meson-exchange models
of the $YN$ interaction. 

For example, the Nijmegen group \cite{MRS89,Rij99,Rij06} views this
interaction as being generated by genuine scalar-meson exchange. In their models 
NSC~\cite{MRS89}, NSC97 \cite{Rij99}, and ESC04 \cite{Rij06} a scalar
SU(3) nonet is exchanged --- namely, two isospin-0 mesons (besides the $\epsilon(760)$,
the $S^*(975)$ ($f_0(980)$) in model NSC (NSC97, ESC04)), an isospin-1 meson 
($\delta$ or $a_0(980)$) and an isospin-1/2 strange meson $\kappa$ with a mass of 1000 MeV.
A genuine scalar SU(3) nonet is also present in the so-called Ehime
potential \cite{Tom01}, where besides the $S^*(975)$ and $\delta$ (or $a_0(980)$)
the $f_0(1581)$ and the $K_0^*(1429)$ are included. In addition the model incorporates
two effective scalar-meson exchanges, $\sigma (484)$ and $\kappa (839)$, that stand
for $(\pi\pi)_{I=0}$ and $(K\pi)_{I=1/2}$ correlations but are treated
phenomenologically. 
In the older $YN$ models of the J\"ulich group \cite{Hol89,Reu94} a $\sigma$ (with
a mass of $\approx$ 550 MeV) is included which is viewed as arising from 
correlated $\pi\pi$ exchange.
In practice, however, the coupling strength of this fictitious $\sigma$ to the
baryons is treated as a free parameter and fitted to the data - a rather 
unsatisfactory feature of those models. 

In the new meson-exchange $YN$ potential presented recently by the J\"ulich group 
a different strategy is followed. Here, indeed, a microscopic model of 
correlated $\pi\pi$ and $K\anti{K}$ exchange is utilized to fix the contributions 
in the scalar-isoscalar ($\sigma$) and vector-isovector ($\rho$) channels. The
basic steps in evaluating these contributions are outlined in the next 
subsection. 
Besides correlated $\pi\pi$ and $K\anti{K}$ exchange the new $YN$ model incorporates 
also the standard one-boson exchange contributions of the lowest
pseudoscalar and vector meson multiplets with coupling constants determined
by SU(3) symmetry relations (\ref{eq:3.9}). The so-called $F/(F+D)$ ratios, cf. Sect. \ref{chap:2.3}, 
are fixed to $\alpha = 0.4$ ($\alpha = 1$) for the pseudoscalar (vector) meson
multiplets by invoking SU(6) symmetry. 
 
Let us mention for completeness that in meson-exchange models usually the
Lippmann-Schwinger equation is not regularized by introducing a regulator 
function of the form (\ref{regulator}) as in the EFT approach.
For example, in case of the $YN$ models of the J\"ulich group \cite{Hol89,Reu94,Hai05} 
convergence of the Lippmann-Schwinger equation is achieved by supplementing the 
interaction with form factors for each meson-baryon-baryon ($xBB'$) vertex, cf. 
Sect. 2.3.3 of Ref.~\cite{Hol89} for details. Those form factors are meant to take
into account the extended hadron structure and are parametrized in the
conventional monopole or dipole form, for example
$F_{xBB'}({\bf{k}}^2) = (\Lambda_{xBB'}^2 - m_x^2)/(\Lambda_{xBB'}^2 + {\bf{ k}}^2)$,
where {$\bf{k}$} is the momentum transfer defined in Eq.~(\ref{eq:3.11}),
$m_x$ is the mass of the exchanged meson and $\Lambda_{xBB'}$ is
the so-called cut--off mass. 

\subsection{Model for correlated $\pi\pi + K\anti{K}$ exchange}

The explicit derivation of the baryon-baryon interaction based on 
correlated $\pi\pi + K\anti{K}$ exchange is quite involved and we refer the
interested reader to the work of Reuber et al. \cite{Reu96} for the 
full details. Here we only describe briefly the principal steps of 
the derivation of the correlated $\pi\pi + K\anti{K}$ exchange 
potentials for the baryon-baryon amplitudes in the scalar-isoscalar 
($\sigma$) and vector-isovector ($\rho$) channels.
  
\begin{figure}[t]
\centering
\resizebox{0.5\textwidth}{!}{\includegraphics[4cm,2cm][16cm,16cm]{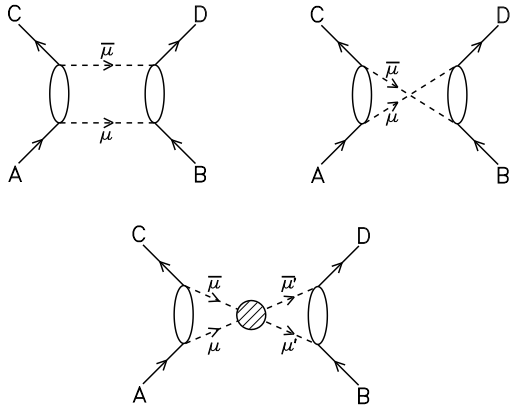}}
\caption{Two-pion and two-kaon exchange in the baryon-baryon process $A+B\to C+D$.
The unshaded ellipse denotes the direct coupling of the two pseudoscalar mesons
$\mu\bar\mu = \pi\pi, \, K\anti{K}, \, \anti{K}K$ to the baryons without any
correlation effects. The shaded circle in the lower diagram for the correlated
exchange stands for the full off-shell amplitude of the process $\mu\bar\mu \to 
\mu'\bar\mu'$.} 
\label{fig1_1}
\end{figure}

Based on a $\pi\pi - K\anti{K}$ amplitude the evaluation of the correlated 
$\pi\pi$ exchange process for the baryon-baryon reaction $A+B \to C+D$, 
cf. the cartoon in Fig.~\ref{fig1_1}, can be done in two steps.
Firstly the $A\overline{C} \rightarrow \pi\pi, K\anti{K}$ amplitude
is determined in the pseudophysical region 
and then dispersion theory and unitarity are applied to connect this
amplitude with the corresponding physical amplitudes in the $A+B \to C+D$ 
channel. In our concrete case $A$, $B$, etc. can be any combination of
the baryons $N$, $\Lambda$, $\Sigma$, or $\Xi$. 

The Born terms for the transitions $A\overline{C} \rightarrow \pi\pi, K\anti{K}$
include contributions from baryon exchange as well as
$\rho$-pole diagrams (cf. Ref.~\cite{Jan95}).
The correlations between the two pseudoscalar mesons are taken into
account by means of a coupled channel ($\pi\pi$, $K\bar{K}$) model
\cite{Jan95,Loh90} generated from $s$- and $t$-channel meson exchange
Born terms.
This model describes the empirical $\pi\pi$ phase shifts over a large
energy range from threshold up to 1.3 GeV. 
The amplitudes for the $A\bar{C} \rightarrow \pi\pi$, $K\bar{K}$
transitions in the pseudophysical region are then obtained by solving
a covariant scattering equation with full inclusion of the $\pi\pi$ - $K\bar{K}$
correlations. The parameters of the $A\bar{C} \rightarrow \pi\pi$, $K\bar{K}$ 
model, which are interrelated through SU(3) symmetry, are determined by fitting
to the quasiempirical $N\bar{N} \rightarrow \pi\pi$ amplitudes in the
pseudophysical region, $t \leq 4 m^2_\pi$ \cite{Reu96}, obtained by
analytic continuation of the empirical $\pi N$ and $\pi\pi$ data.

Assuming analyticity for the amplitudes dispersion relations
can be formulated for the baryon-baryon amplitudes, which connect
physical amplitudes in the $s$-channel with singularities and
discontinuities of these amplitudes in the pseudophysical region of
the $t$-channel processes for the $J^P = 0^+$ ($\sigma$) and $1^-$ ($\rho$)
channel:
\begin{equation}
V^{(0^+,1^-)}_{A,B \to C,D}(t) \propto \int_{4m^2_\pi}^\infty
dt'
{ {\rm Im} V^{(0^+,1^-)}_{A,\overline{C} \to \overline{B},D}(t') \over t'-t}, \ \  t < 0 .
\label{dispersion}
\end{equation}
Via unitarity relations the singularity structure of the baryon-baryon
amplitudes for $\pi\pi$ and $K\anti{K}$ exchange are fixed by and
can be written as products of the
$A\anti{C}\to\pi\pi,\,K\anti{K}$ amplitudes
\begin{equation}
{\rm Im} V^{(0^+,1^-)}_{A,\overline{C} \to \overline{B},D}(t') \propto
\sum_{\alpha = \pi\pi, K\anti{K}} T^{*,(0^+,1^-)}_{A,\overline{C} \to \alpha}
\, T^{(0^+,1^-)}_{\overline{B},D \to \alpha}.
\label{unitarity}
\end{equation}
Thus,
from the $A\anti{C} \rightarrow \pi\pi, \, K\anti{K} $ amplitudes the
spectral functions can be calculated
\begin{equation}
\rho^{(0^+,1^-)}_{A,B \to C,D}(t') \propto
\sum_{\alpha = \pi\pi, K\anti{K}} T^{*,(0^+,1^-)}_{A,\bar{C} \to \alpha}
\, T^{(0^+,1^-)}_{\bar{B},D \to \alpha}
\label{spectral}
\end{equation}
which are then inserted into dispersion integrals to
obtain the (on-shell) baryon-baryon interaction:
\begin{equation}
V^{(0^+,1^-)}_{A,B \to C,D}(t) \propto \int_{4m^2_\pi}^\infty
dt'
{\rho^{(0^+,1^-)}_{A,B \to C,D}(t') \over t'-t}, \ \  t < 0 .
\label{potential}
\end{equation}

The spectral function (\ref{spectral}) for the ($0^+$) $\sigma$-channel
has only one component but the one for the ($1^-$) $\rho$-channel
consists of four linearly independent components, which reflects the 
more complicated spin structure of this channel \cite{Reu96}. Note that the 
amplitudes in Eq.~(\ref{dispersion}) still contain the uncorrelated 
(upper diagrams in Fig.~\ref{fig1_1}), as well as the correlated pieces
(lower diagram). Thus, in order to obtain the contribution of the truly 
correlated
$\pi\pi$ and $K\anti{K}$ exchange one must eliminate the former from
the spectral function.
This is done by calculating the spectral function generated by
the Born term and subtracting it from the total spectral function:
\begin{equation}
\rho^{(0^+,1^-)} \longrightarrow \rho^{(0^+,1^-)} -
\rho^{(0^+,1^-)}_{\rm Born} .
\end{equation}
We should mention that the uncorrelated contributions are included 
too. But they are generated automatically by solving the scattering 
equation (\ref{eq:ls}) for the interaction potential. 

Finally, let us mention
that the spectral functions characterize both the strength
and range of the interaction.
Clearly, for sharp mass exchanges the spectral function becomes
a $\delta$-function at the appropriate mass.

For convenience in the concrete calculations the potential due
to correlated $\pi\pi/K\anti{K}$ exchange is parametrized in 
terms of effective coupling strengths of (sharp mass) 
$\sigma$ and $\rho$ exchanges.
The interaction resulting from the exchange of a
$\sigma$ meson with mass $m_\sigma$ between two $J^P=1/2^+$
baryons $A$ and $B$ has the structure:
\begin{equation}
V^{\sigma}_{A,B \to A,B}(t) \ = \ g_{AA\sigma} g_{BB\sigma}
{F^2_\sigma (t) \over t - m^2_\sigma} ,
\label{formd}
\end{equation}
where a form factor $F_\sigma(t)$ is applied at each vertex,
taking into account the fact that the exchanged $\sigma$ meson is
not on its mass shell.
The correlated potential as given in Eq.~(\ref{dispersion}) can now be
parameterized in terms of $t$-dependent strength functions
$G_{AB \to AB}(t)$, so that for the $\sigma$ case:
\begin{equation}
V^{(0^+)}_{A,B \to A,B}(t) =
G^{\sigma}_{AB \to AB}(t) F^2_\sigma(t) {1 \over t - m^2_\sigma}.
\label{sigma}
\end{equation}
The effective coupling constants are then defined as 
\begin{equation}
g_{AA\sigma}g_{BB\sigma} \quad\longrightarrow \quad G_{AB\to
AB}^\sigma (t)= {(t-m_\sigma^2)\over\pi F^2_\sigma(t)}
\int_{4m_\pi^2}^{\infty} {\rho^{(0^+)}_{AB \to AB}(t') \over t'-t} dt' .
\label{effccsig}
\end{equation}
In the concrete application one varies $m_\sigma^2$ in order to achieve 
that $G_{AB\to AB}^\sigma (t) \approx G_{AB\to AB}^\sigma$, i.e. 
that $G_{AB\to AB}^\sigma$ is indeed practically a constant. 
The form factor is parameterized by 
\begin{equation}
F_\sigma (t) = {\Lambda ^2_\sigma \over
\Lambda ^2_\sigma - t} \ ,
\label{form}
\end{equation}
with a cut--off mass $\Lambda_\sigma$ assumed to be the same
for both vertices. This form guarantees that the on-shell behaviour of the 
potential (which is fully determined by the dispersion integral) is not 
modified strongly as long as the energy is not too high.

Similar relations can be also derived for the correlated exchange
in the isovector-vector channel \cite{Reu96}, which in this case
will involve vector as well as tensor coupling pieces.

\subsection{Other ingredients of the J\"ulich meson-exchange hyperon-nucleon model}

Besides the correlated $\pi\pi$ and $K\bar{K}$ exchange the new $YN$ model of the
J\"ulich group takes into account exchange diagrams
involving the well-established lowest lying pseudoscalar and vector
meson SU(3) octets. Following the philosophy of the original J\"ulich
$YN$ potential \cite{Hol89} the coupling constants in the pseudoscalar
sector are fixed by strict SU(6) symmetry. In any case, this is
also required for being consistent with the model of correlated $\pi\pi$
and $K\anti{K}$ exchange. The cut--off masses of the
form factors (cf. discussion at the beginning of Sect. \ref{chap:3})
belonging to the $NN$ vertices are taken over from the
full Bonn $NN$ potential. The cut--off masses at the strange vertices
are considered as open parameters though, in practice, their values
are kept as close as possible to those found for the $NN$ vertices.

In addition there are some other new ingredients in the present $YN$ 
model as compared to the earlier J\"ulich models \cite{Hol89,Reu94}. 
First of all, we now take into account contributions from 
(scalar-isovector) $a_0(980)$ exchange.
The $a_0$ meson is present in the original Bonn $NN$ potential
\cite{MHE}, and for consistency should also be included in the $YN$
model. Secondly, we consider the exchange of a strange scalar meson, 
the $\kappa$, with mass $\sim 1000$~MeV.
Let us emphasize, however, that like in case of the $\sigma$ meson
these particles are not viewed as being members of a scalar meson SU(3)
multiplet, but rather as representations
of strong meson-meson correlations in the scalar--isovector
($\pi\eta$--$K\anti{K}$) \cite{Jan95} and scalar--isospin-1/2 ($\pi K$)
channels \cite{Loh90}, respectively.
In principle, their contributions can also be evaluated along the lines
of Ref.~\cite{Reu96}, however, for simplicity in the present model they
are effectively parameterized by one-boson-exchange diagrams with the
appropriate quantum numbers assuming the coupling constants to be free
parameters. 
Furthermore, the new model contains the exchange of an $\omega '$ with a mass 
of $m_{\omega '}$ = 1120 MeV considered to be an effective parametrization 
of short-range contributions from correlated $\pi-\rho$ exchange 
\cite{Jan96} in the vector-isoscalar sector. Its inclusion allows to keep the 
coupling constants of the genuine $\omega$(782) meson to the baryons 
in line with their SU(3) values, cf. the discussion in Ref. \cite{Hai05}.
In the spirit of
the EFT approach, we have also considered a version of the $YN$ potential
in \cite{Hai05} where the $\kappa$ exchange was substituted by a 
local contact interaction.

\begin{figure}[t]
\centering
\resizebox{\textwidth}{!}{\includegraphics*[2cm,16.5cm][13cm,26cm]{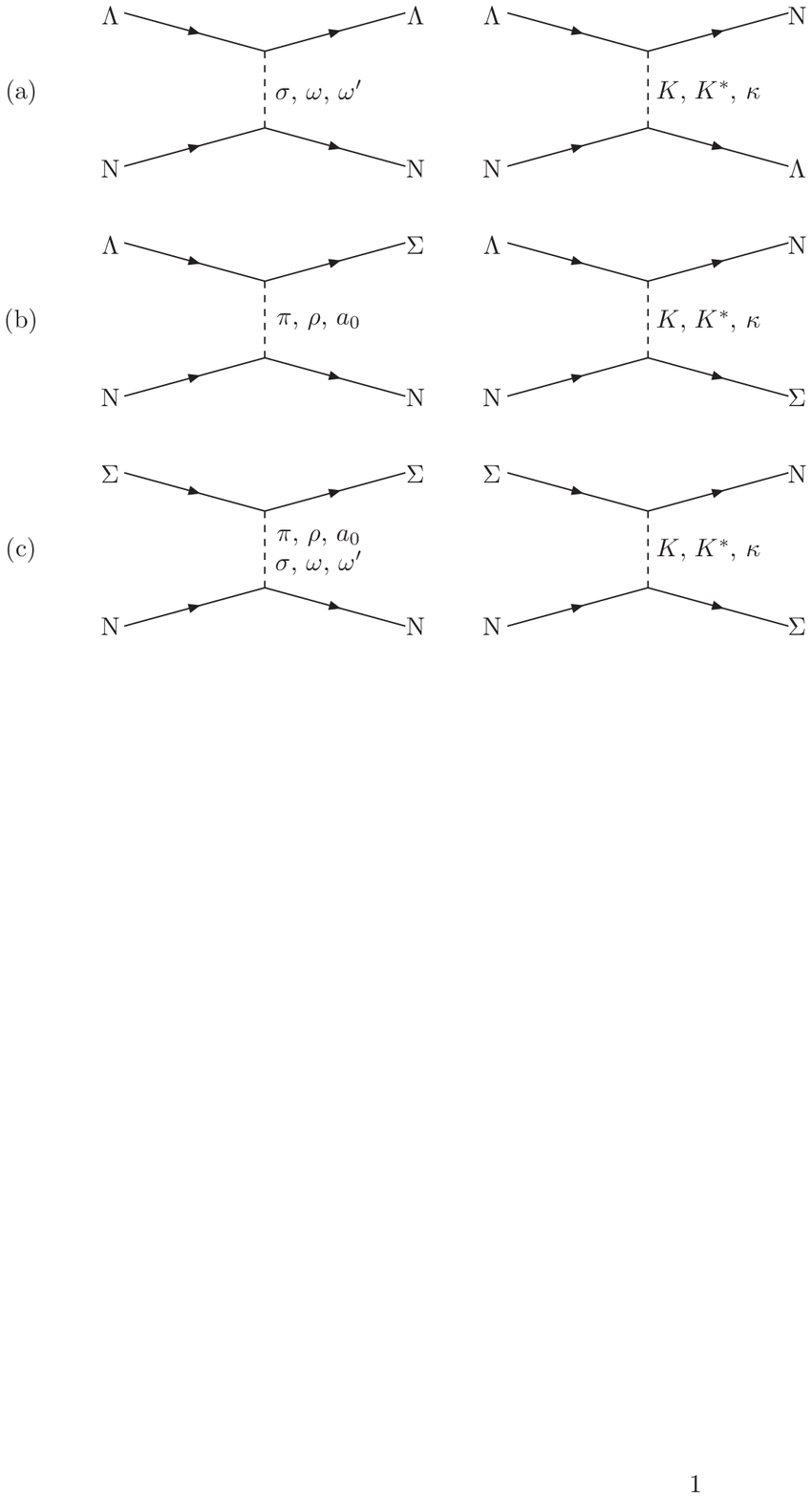}}
\caption{Contributions to the meson-exchange $YN$ model \cite{Hai05} in the
$\Lambda N$ and $\Sigma N$ channels and in the
$\Lambda N \to \Sigma N$ transition. Note that only
$\pi$, $K$, $\omega$, and $K^*$ exchange are considered as being
due to genuine SU(3) mesons. The other contributions are
either fixed from correlated $\pi\pi$ and $K\anti{K}$ exchange
($\sigma$, $\rho$) or are viewed as an effective parametrization
of meson-meson correlations ($a_0$, $\kappa$, $\omega$')
in the corresponding spin-isospin channels.
}
\label{figyn}
\end{figure}

Thus we have the following scenario: The long- and intermediate-range part
of the new meson-exchange $YN$ interaction model is completely
determined by SU(6) constraints (for the pseudoscalar and, in general,
also for the vector mesons) and by correlated
$\pi\pi$ and $K\anti{K}$ exchange. The short-range part is viewed as
being due to correlated meson-meson exchanges but in practice is
parametrized phenomenologically in terms of one-boson-exchange
contributions in specific spin-isospin channels. In particular,
no SU(3) relations are imposed on the short-range part. This
assumption is based on our observation that the contributions in the
$\rho$ exchange channel as they result from
correlated $\pi\pi$ and $K\bar{K}$ no longer fulfill strict
SU(3) relations \cite{Reu96}, but it also acknowledges
the fact that at present there is no general agreement about who are
the actual members of the lowest-lying scalar meson SU(3) multiplet,
as already mentioned above. 
A graphical representation of all meson-exchange contributions that are
included in the new $YN$ model is given in Fig. \ref{figyn}.

\section{Results and discussion}
\label{chap:6}

\begin{figure}
\resizebox{\textwidth}{!}{%
  \includegraphics*[2.0cm,16.9cm][10.75cm,27cm]{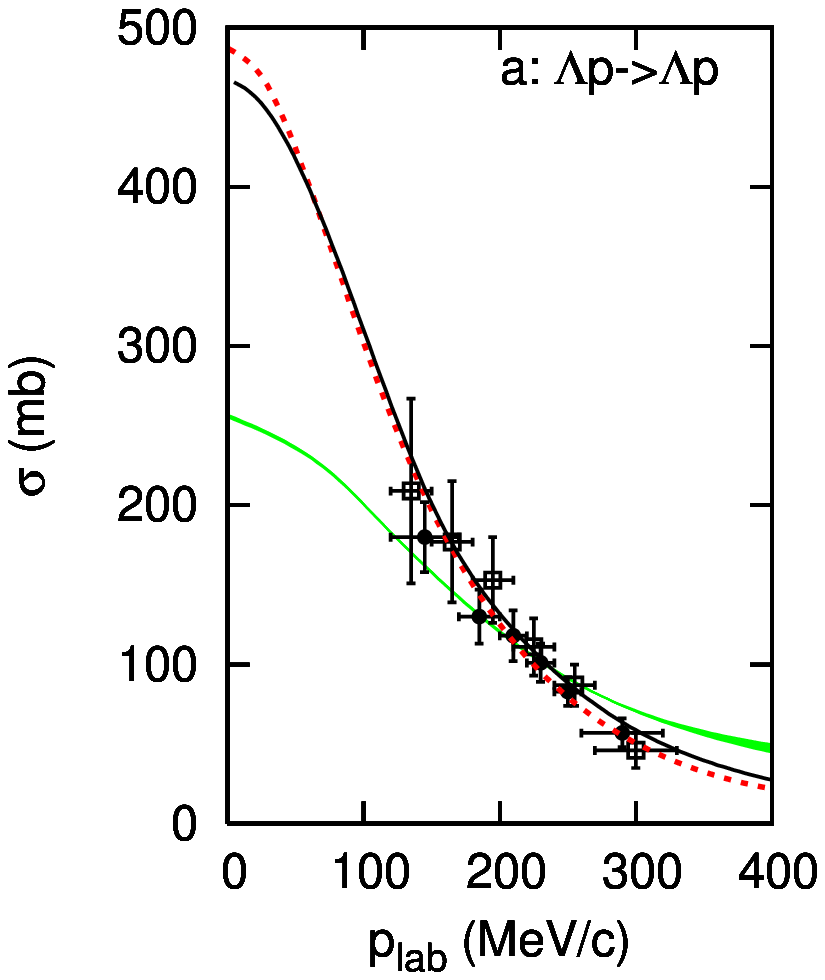}
  \includegraphics*[2.0cm,16.9cm][10.75cm,27cm]{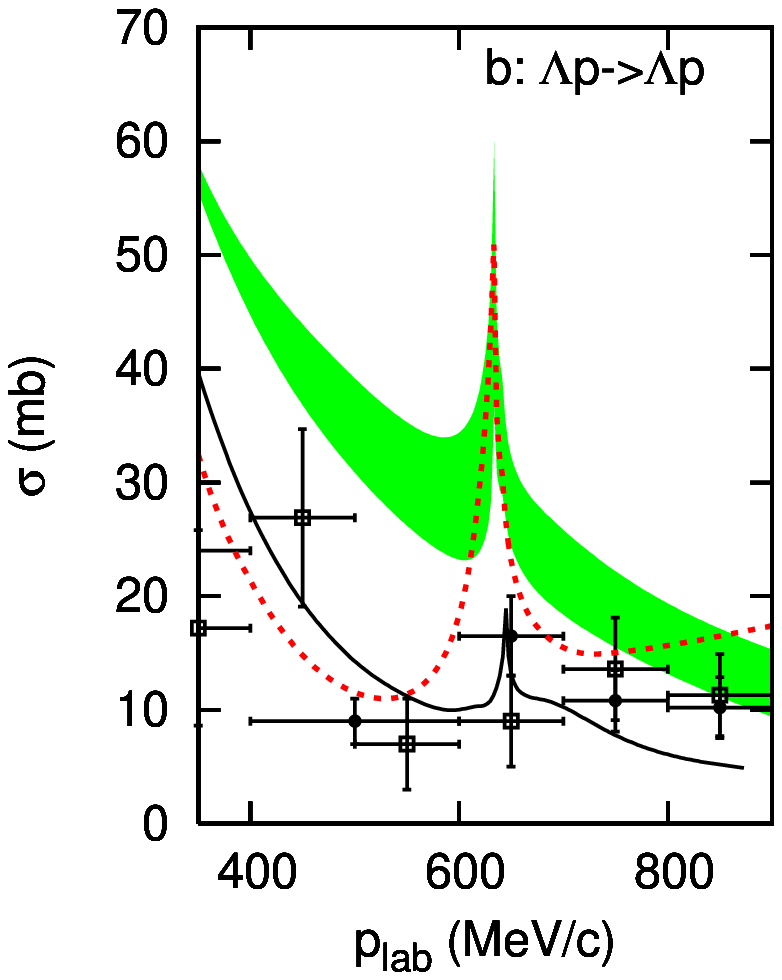}
  \includegraphics*[2.0cm,16.9cm][10.75cm,27cm]{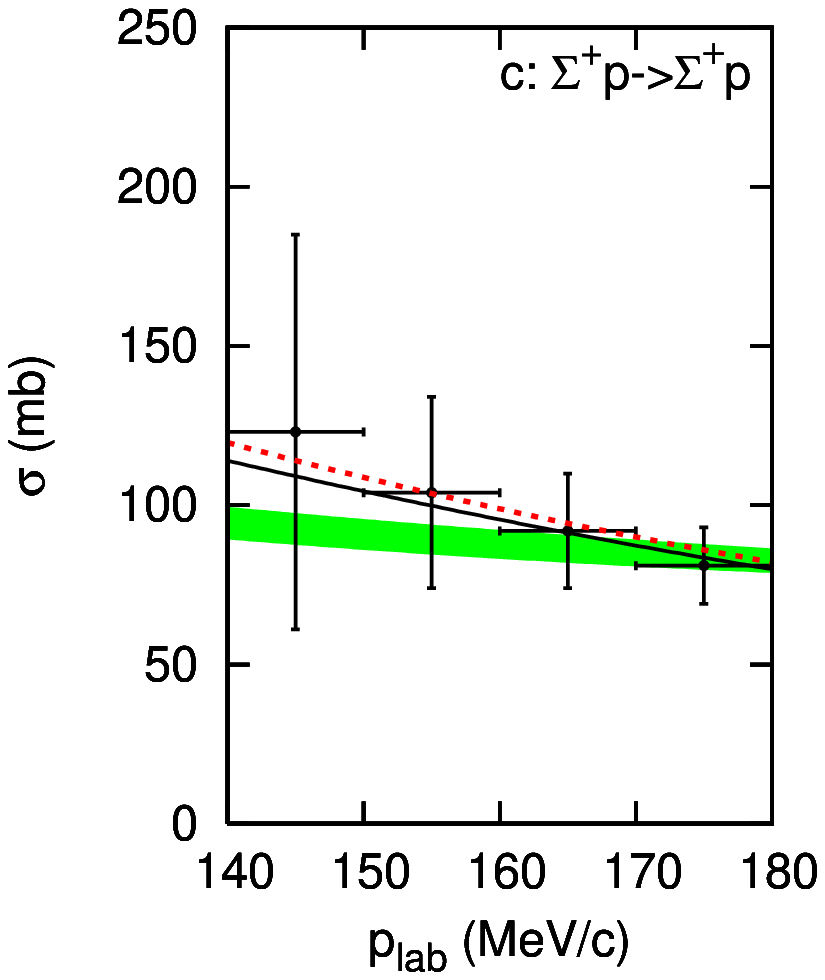}
}
\hfill \break
\resizebox{\textwidth}{!}{%
  \includegraphics*[2.0cm,16.9cm][10.75cm,27cm]{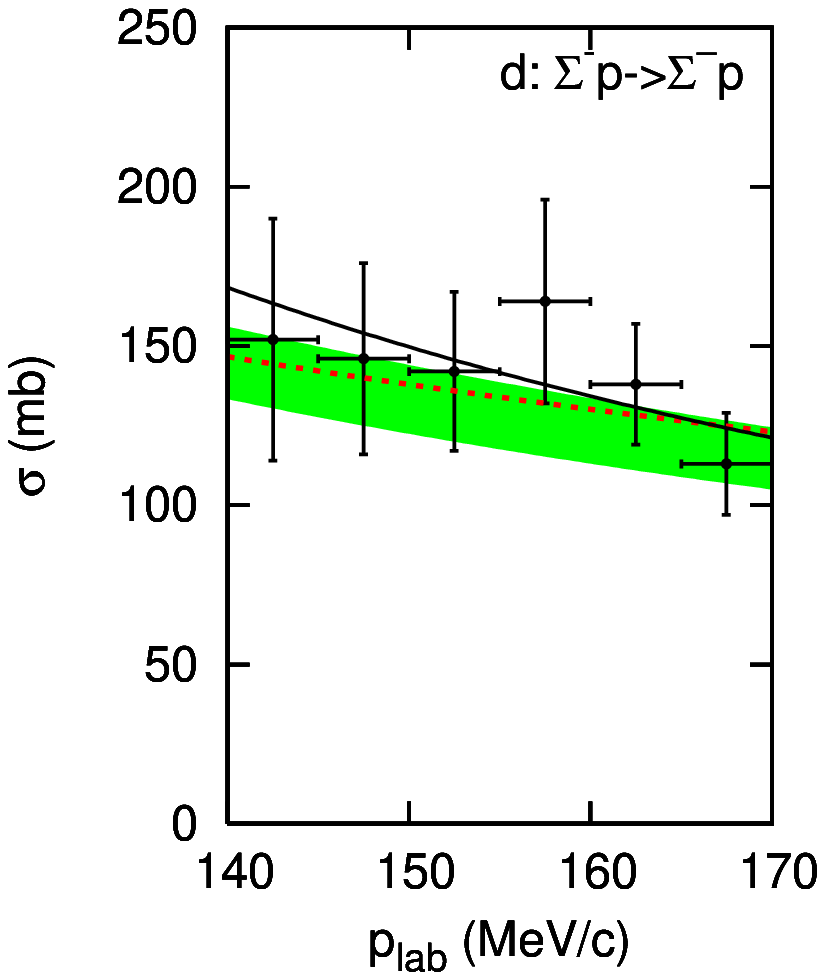}
  \includegraphics*[2.0cm,16.9cm][10.75cm,27cm]{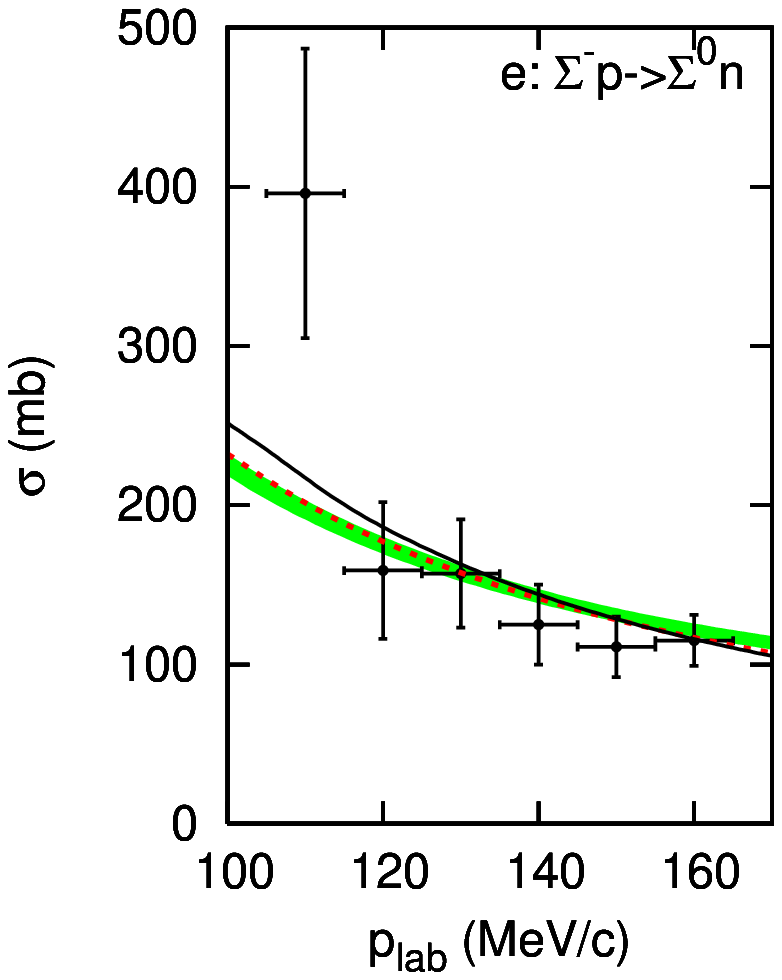}
  \includegraphics*[2.0cm,16.9cm][10.75cm,27cm]{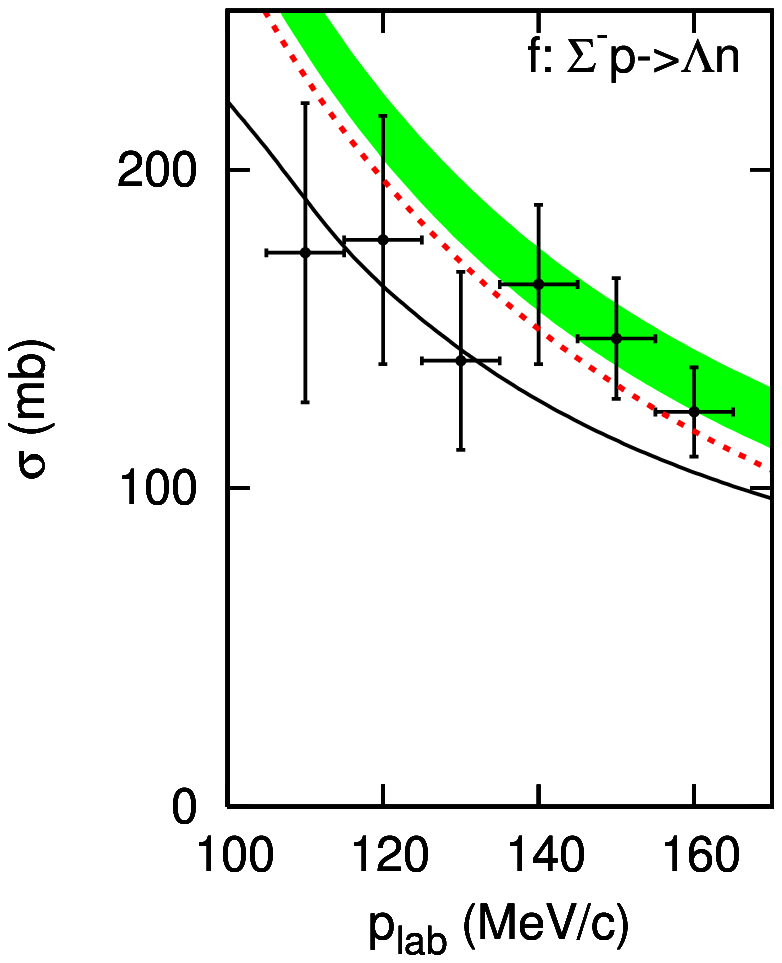}
}
\caption{''Total'' cross section $\sigma$ (as defined in Eq.~(\ref{eq:sigtot}))
as a function of $p_{{\rm lab}}$. The experimental cross sections in (a) are taken from Refs.~\cite{Sec68} (open squares) and ~\cite{Ale68} (filled circles), in (b) from Refs.~\cite{Kad71} (filled circles) and \cite{Hau77} (open squares) , in (c),(d) from \cite{Eis71} and in (e),(f) from \cite{Eng66}. The shaded band is the chiral EFT result for $\Lambda = 550,...,700$ MeV \cite{Pol06}, 
the solid curve is the J{\"u}lich '04 model \cite{Hai05}, 
and the dashed curve is the Nijmegen NSC97f potential \cite{Rij99}.}
\label{fig:6.0}
\end{figure}

In Fig. \ref{fig:6.0} we confront the results obtained from
our $YN$ interactions with the $\Lambda p$, $\Sigma^+p$, $\Sigma^-p$,
$\Sigma^-p \to \Sigma^0n$, and $\Sigma^-p \to \Lambda n$
data used in the fitting procedure. 
Here the solid curves correspond to the J\"ulich '04 meson-exchange model and 
the shaded band represents the results of the chiral EFT for the considered cut--off region. 
For reasons of comparison we also include the results of one of the meson-exchange models
(NSC97f) of the Nijmegen group (dashed line) \cite{Rij99}. 
A detailed comparison between the experimental scattering data considered and the values 
found in the fitting procedure for the EFT interaction (for $\Lambda = 600$ MeV) is given 
in Table \ref{tab:6.1a}.
\begin{table}[t]
\caption{Comparison between the 35 experimental $YN$ data and the results for the EFT interaction
  for the cut--off $\Lambda = 600$ MeV. Momenta are in units of MeV and 
cross sections in mb. The achieved $\chi^2$ is given for each reaction channel separately. 
}
\label{tab:6.1a}
\vspace{0.2cm}
\centering
\begin{tabular}{|rrr|rrr|rrr|}
\hline
\multicolumn{3}{|c|}{$\Lambda p \rightarrow \Lambda p\;\;$ $\chi^2 = 7.5$}& \multicolumn{3}{c|}{$\Lambda p \rightarrow \Lambda p\;\;$ $\chi^2 = 4.9$}& \multicolumn{3}{c|}{$\Sigma^- p \rightarrow \Lambda n\;\;$ $\chi^2 = 5.5$}\\
$p_{{\rm lab}}^{\Lambda}$& $\sigma_{{\rm exp}}$\cite{Sec68}& $\sigma_{{\rm the}}$& $p_{{\rm lab}}^{\Lambda}$& $\sigma_{{\rm exp}}$\cite{Ale68}& $\sigma_{{\rm the}}$& $p_{{\rm lab}}^{\Sigma^-}$& $\sigma_{{\rm exp}}$\cite{Eng66}& $\sigma_{{\rm the}}$ \\
\hline
135 &209$\pm$58 &170.0 &145 &180$\pm$22 &161.6 &110 &174$\pm$47 &244.2 \\
165 &177$\pm$38 &145.4 &185 &130$\pm$17 &130.4 &120 &178$\pm$39 &210.0 \\
195 &153$\pm$27 &123.5 &210 &118$\pm$16 &113.7 &130 &140$\pm$28 &183.0 \\
225 &111$\pm$18 &104.7 &230 &101$\pm$12 &101.9 &140 &164$\pm$25 &161.4 \\
255 &87 $\pm$13 &89.1  &250 &83 $\pm$13  &91.5  &150 &147$\pm$19 &143.9 \\
300 &46 $\pm$11 &70.6  &290 &57 $\pm$9  &74.3  &160 &124$\pm$14 &129.5 \\
\hline
\hline
\multicolumn{3}{|c|}{$\Sigma^+ p \rightarrow \Sigma^+ p\;\;$ $\chi^2 = 0.6$}& \multicolumn{3}{c|}{$\Sigma^- p \rightarrow \Sigma^- p\;\;$ $\chi^2 = 2.4$}& \multicolumn{3}{c|}{$\Sigma^- p \rightarrow \Sigma^0 n\;\;$ $\chi^2 = 7.0$}\\
$p_{{\rm lab}}^{\Sigma^+}$& $\sigma_{{\rm exp}}$\cite{Eis71}& $\sigma_{{\rm the}}$& $p_{{\rm lab}}^{\Sigma^-}$& $\sigma_{{\rm exp}}$\cite{Eis71}& $\sigma_{{\rm the}}$& $p_{{\rm lab}}^{\Sigma^-}$& $\sigma_{{\rm exp}}$\cite{Eng66}& $\sigma_{{\rm the}}$ \\
\hline
145 &123$\pm$62 &96.7 &142.5 &152$\pm$38 &143.4 &110 &396$\pm$91 &200.0 \\
155 &104$\pm$30 &93.0 &147.5 &146$\pm$30 &137.5 &120 &159$\pm$43 &177.4 \\
165 &92 $\pm$18 &89.6 &152.5 &142$\pm$25 &131.9 &130 &157$\pm$34 &159.3 \\
175 &81 $\pm$12 &86.7 &157.5 &164$\pm$32 &126.8 &140 &125$\pm$25 &144.7 \\
    &           &     &162.5 &138$\pm$19 &122.1 &150 &111$\pm$19 &132.7 \\
    &           &     &167.5 &113$\pm$16 &117.6 &160 &115$\pm$16 &122.7 \\
\hline
\hline
\multicolumn{3}{|c}{$r^{{\rm exp}}_R = 0.468\pm 0.010$}& \multicolumn{3}{c}{$r^{{\rm the}}_R = 0.475$}& \multicolumn{3}{c|}{$\chi^2 = 0.5$}\\
\hline
\end{tabular}
\end{table}
The differential cross sections are calculated in the usual way using the 
partial wave amplitudes, for details we refer to \cite{Hol89,Fra80}. 
The total cross sections are found by simply 
integrating the differential cross sections, except for the 
$\Sigma^+ p\to \Sigma^+ p$ and $\Sigma^- p\rightarrow \Sigma^- p$
channels. For those channels 
the experimental total cross sections were obtained via \cite{Eis71}
\begin{eqnarray}\label{eq:sigtot}
\sigma&=&\frac{2}{\cos \theta_{{\rm max}}-\cos \theta_{{\rm min}}}\int_{\cos \theta_{{\rm min}}}^{\cos \theta_{{\rm max}}}\frac{d\sigma(\theta)}{d\cos \theta}d\cos \theta \ ,
\end{eqnarray}
for various values of $\cos \theta_{{\rm min}}$ and $\cos \theta_{{\rm max}}$. Following \cite{Rij99}, we use $\cos \theta_{{\rm min}}=-0.5$ and $\cos \theta_{{\rm max}}=0.5$ in our calculations for the $\Sigma^+ p\rightarrow \Sigma^+ p$ and $\Sigma^- p\rightarrow \Sigma^- p$ cross sections, in order to stay as close as possible 
to the experimental procedure.

\begin{figure}
\resizebox{\textwidth}{!}{%
  \includegraphics*[2.0cm,16.9cm][10.75cm,27cm]{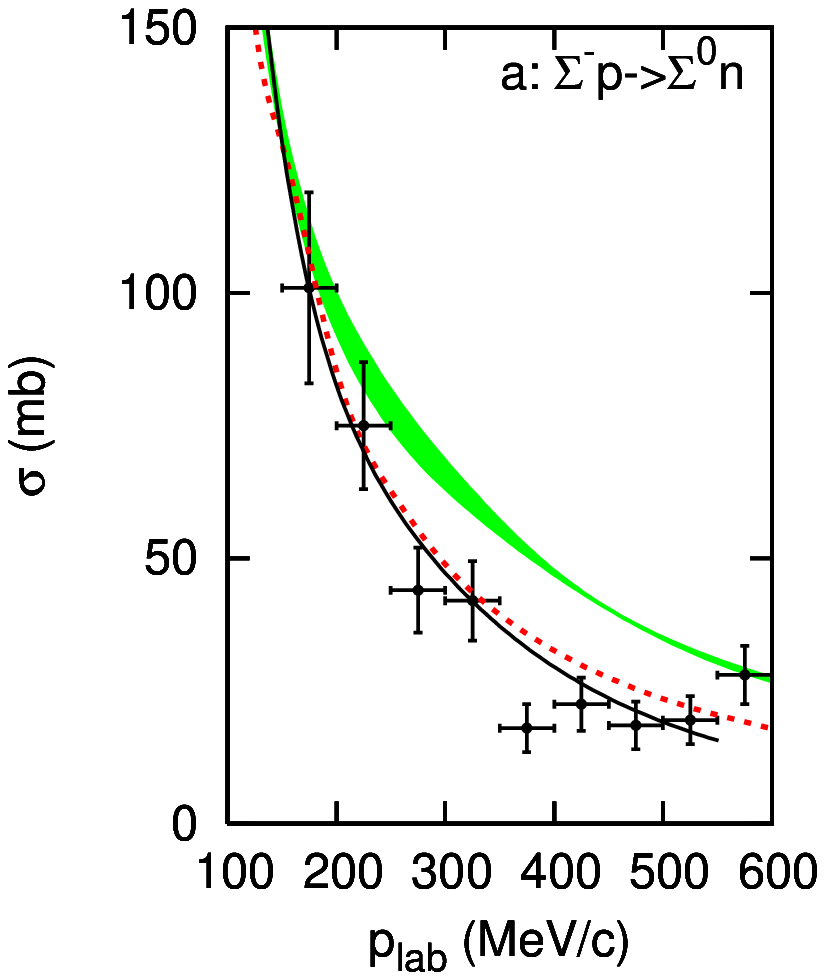}
  \includegraphics*[2.0cm,16.9cm][10.75cm,27cm]{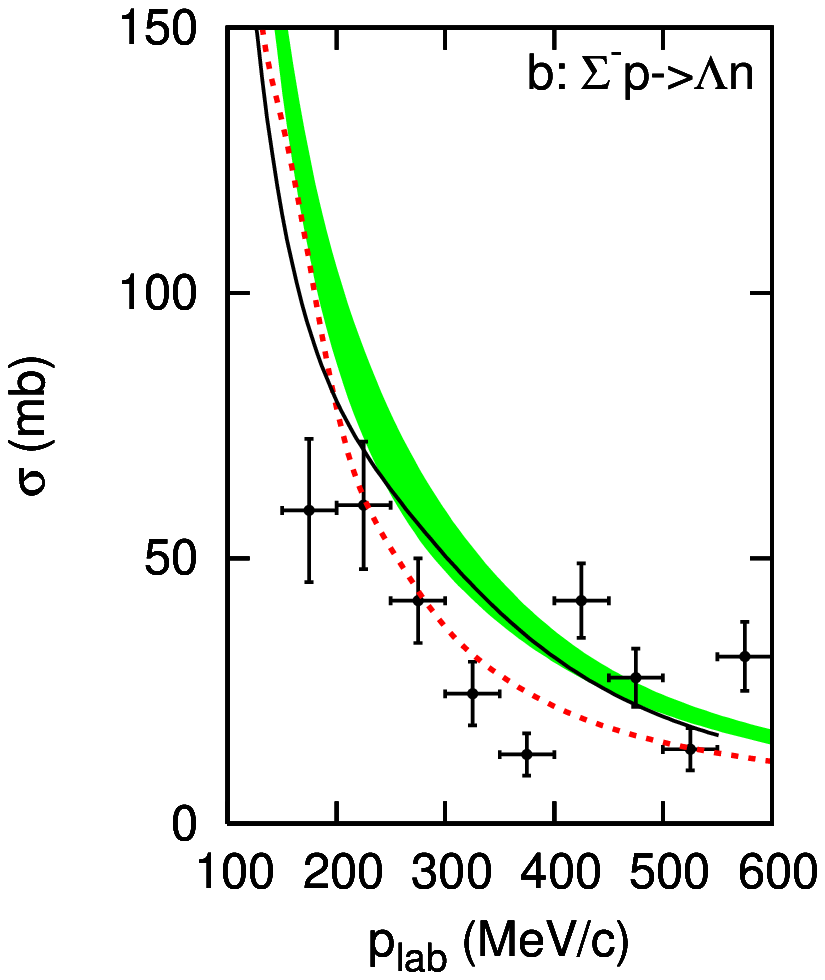}
  \includegraphics*[2.0cm,16.9cm][10.75cm,27cm]{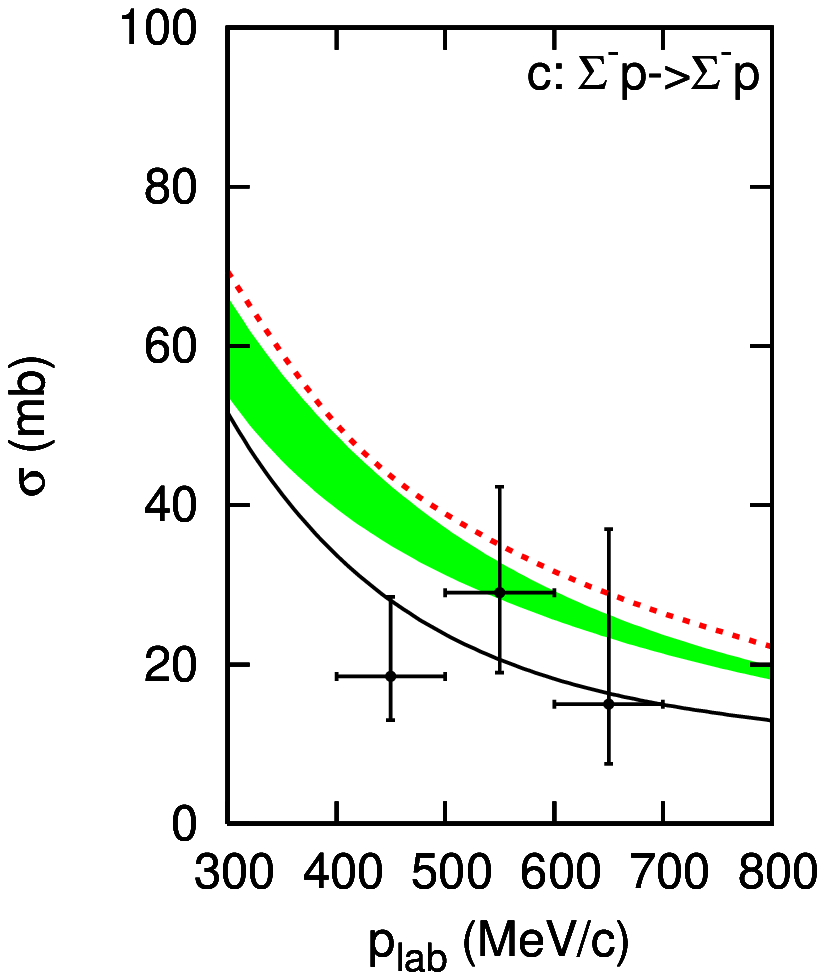}
}
\caption{As in Fig. \ref{fig:6.0}, but now the experimental cross sections in (a),(b) are taken from Refs.~\cite{Ste70} and in (c) from \cite{Kon00}.}
\label{fig:6.1}
\end{figure}

A good description of the low-energy $YN$ scattering data has been obtained with the 
discussed meson-exchange models but also within the EFT approach in the considered 
cut--off region, as is documented in Tables \ref{tab:6.1} 
and \ref{tab:6.1a} and Figs. \ref{fig:6.0} and \ref{fig:6.1}. 

\begin{figure}
\resizebox{\textwidth}{!}{%
  \includegraphics*[2.0cm,16.9cm][10.75cm,27cm]{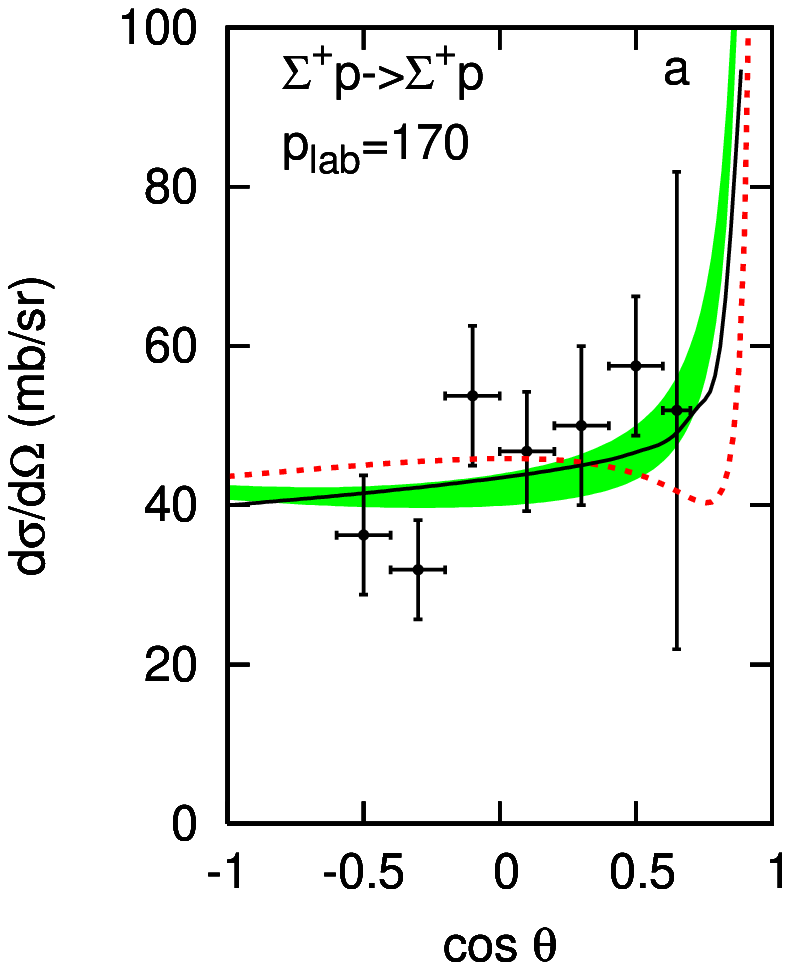}
  \includegraphics*[2.0cm,16.9cm][10.75cm,27cm]{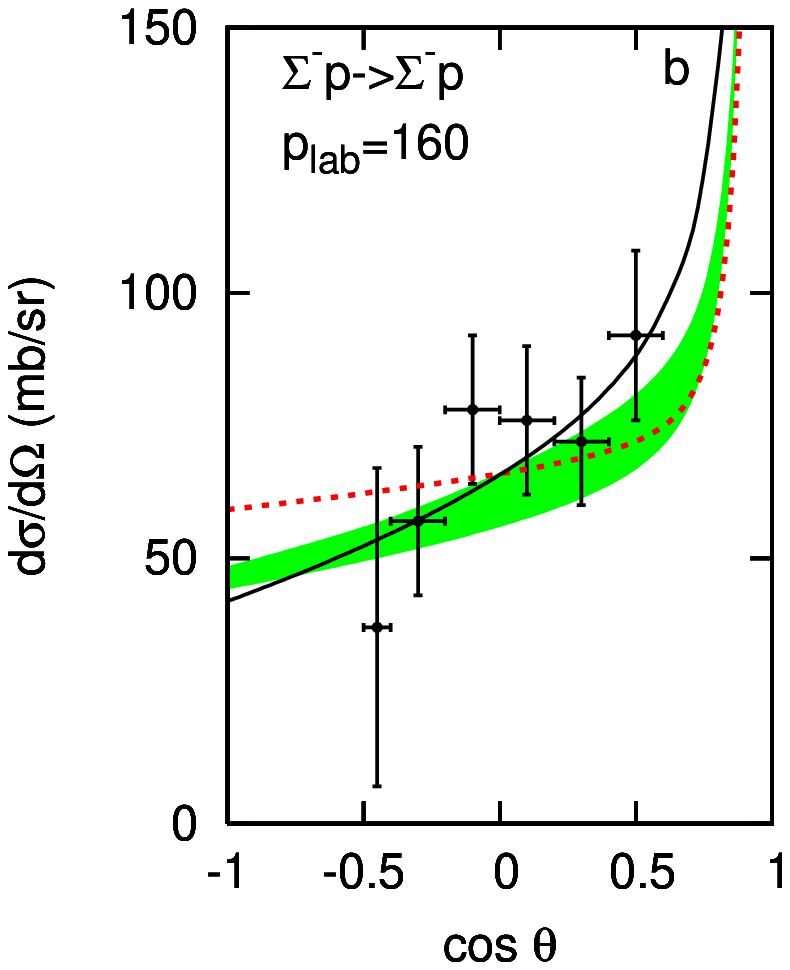}
  \includegraphics*[2.0cm,16.9cm][10.75cm,27cm]{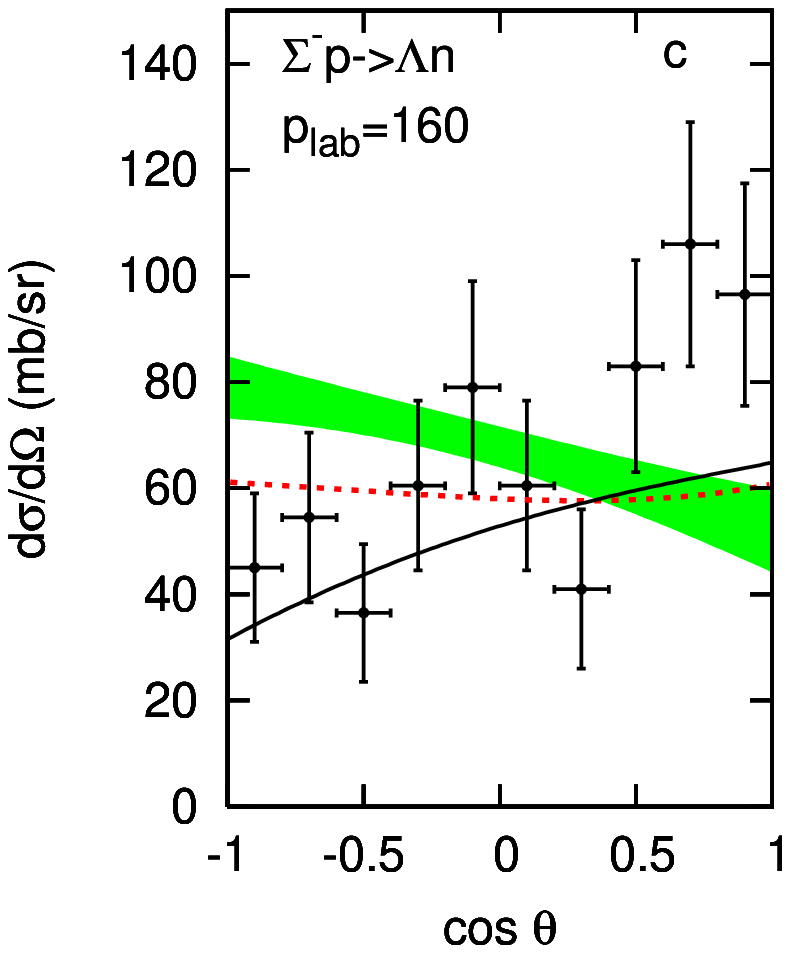}
}
\hfill \break
\resizebox{\textwidth}{!}{%
  \includegraphics*[2.0cm,16.9cm][10.75cm,27cm]{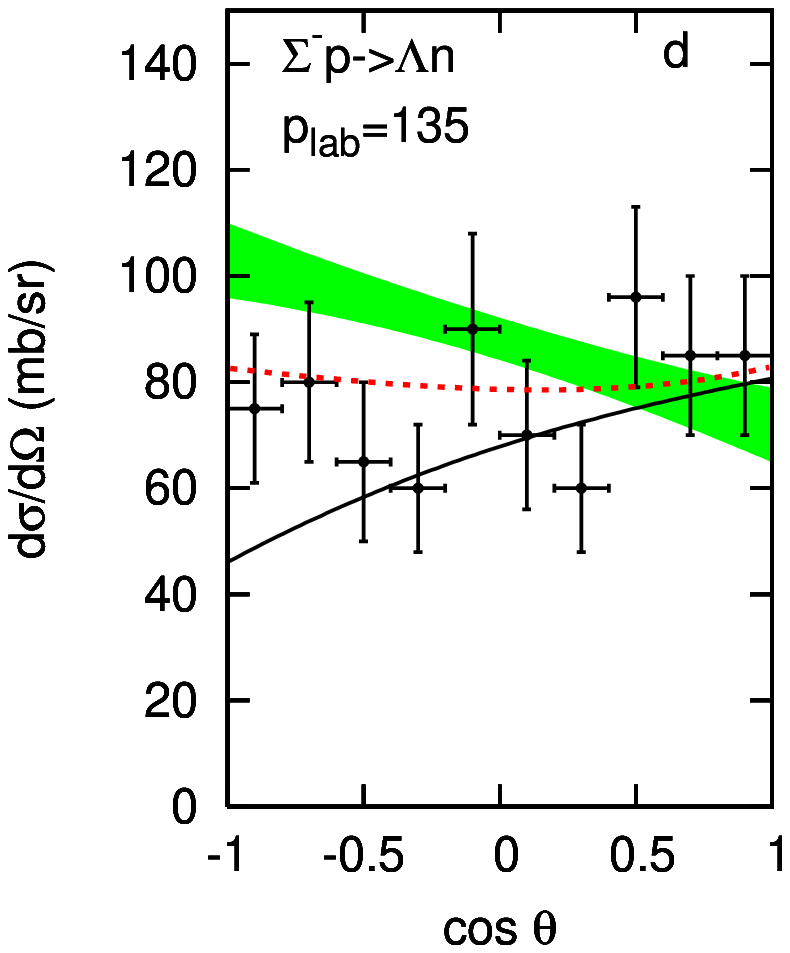}
  \includegraphics*[2.0cm,16.9cm][10.75cm,27cm]{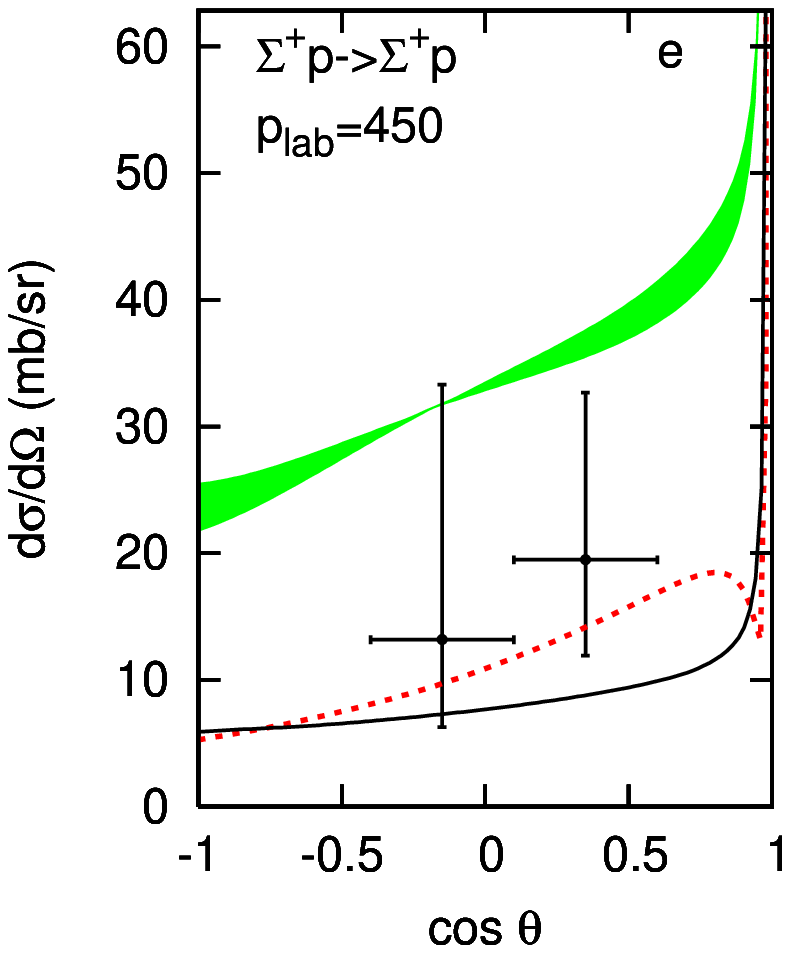}
  \includegraphics*[2.0cm,16.9cm][10.75cm,27cm]{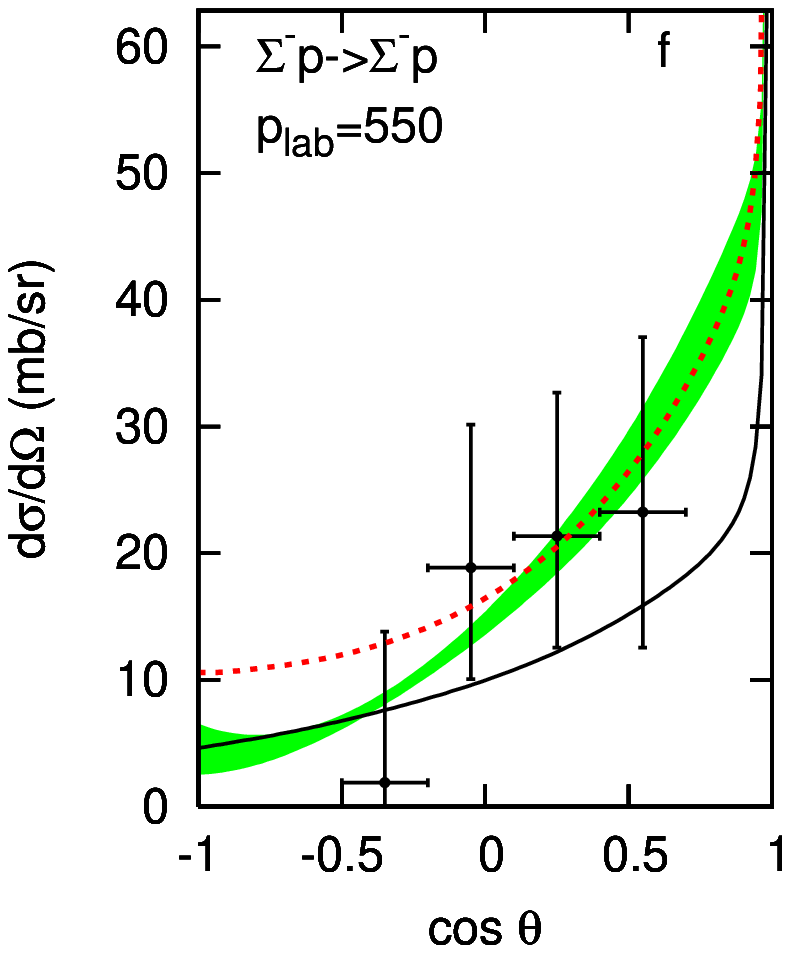}
}
\caption{Differential cross section $d\sigma / d\cos \theta$ as a function of $\cos \theta$, where $\theta$ is the c.m. scattering angle, at various values of $p_{{\rm lab}}$ (MeV/c). The experimental differential cross sections in (a),(b) are taken from \cite{Eis71}, in (c),(d) from \cite{Eng66}, in (e) from \cite{Ahn99} and in (f) from \cite{Kon00}. Same description of curves as in Fig. \ref{fig:6.0}.}
\label{fig:6.2}
\end{figure}

Note that in the low-energy regime the cross sections are mainly given by the $S$-wave contribution, except for 
for the $\Lambda N\rightarrow \Sigma N$ cross section where 
the ${}^3D_1(\Lambda N)\leftrightarrow {}^3S_1(\Sigma N)$ transition provides the main contribution.
Still all partial waves with total angular momentum $J\le 2$ were included in the computation of the
observables. 
The $\Lambda p$ cross sections show a clear cusp at the $\Sigma^+ n$ threshold, see Fig. \ref{fig:6.0}b. 
This cusp is very pronounced for the EFT interaction, peaking at 60 mb, but also in case of the 
Nijmegen NSC97f model. It is hard to see this effect in the experimental data, since it occurs over 
a very narrow energy range. 
In case of the EFT interaction the predicted $\Lambda p$ cross section at higher energies is too large 
(cf. Fig. \ref{fig:6.0}b),
which is related to the problem that some LO phase shifts are too large at higher energies. Note that 
this is also the case for the $NN$ interaction \cite{Epe00a}. In a NLO calculation this problem will 
probably vanish. The differential cross sections at low energies, which have not been taken into 
account in the fitting procedure, are predicted well, see Fig. \ref{fig:6.2}. The results of 
the meson-exchange models and of the chiral EFT are also in good agreement with data for total 
cross sections at higher energy \cite{Ste70,Kon00} which were likewise not included in the 
fitting procedure, as can be seen in Fig. \ref{fig:6.1}.

The $\Lambda p$ and $\Sigma^+p$ scattering lengths and effective ranges are listed in Table~\ref{tab:6.2} 
together with the corresponding hypertriton binding energies (preliminary results of $YNN$ Faddeev 
calculations from \cite{Nog06}).
\begin{table}[h]
\caption{The $YN$ singlet and triplet scattering lengths and effective ranges
  (in fm) and the hypertriton binding energy, $E_B$ (in MeV). 
  The binding energies for the 
  hypertriton (last row) \cite{Nog06} are calculated using 
  the Idaho-N3LO $NN$ potential \cite{Entem:2003ft}. The
  experimental value of the hypertriton binding energy is $-2.354(50)$~MeV
  \cite{Gibson:1995an}.
  We notice that the deuteron binding energy is $-2.224$~MeV. 
}
\label{tab:6.2}
\vspace{0.2cm}
\centering
\begin{tabular}{|c|rrrr|r|r|}
\hline
& \multicolumn{4}{|c|}{EFT '06} & \, {J\"ulich '04}\, & \, {NSC97f \cite{Rij99}} \, \\
$\Lambda$ [MeV]& 550& 600& 650& 700 & & \\
\hline
$a^{\Lambda p}_s$ &$-1.90$ &$-1.91$ &$-1.91$ &$-1.91$ & $-2.56$ & $-2.51$\\
$r^{\Lambda p}_s$  &$1.44$  &$1.40$  &$1.36$  &$1.35$ & $2.75$ & $3.03$\\
$a^{\Lambda p}_t$ &$-1.22$ &$-1.23$ &$-1.23$ &$-1.23$ & $-1.66$ & $-1.75$\\
$r^{\Lambda p}_t$  &$2.05$  &$2.13$  &$2.20$  &$2.27$ & $2.93$ & $3.32$\\
\hline
$a^{\Sigma^+ p}_s$  &$-2.24$ &$-2.32$  &$-2.36$  &$-2.29$ & $-4.71$ & $-4.35$\\
$r^{\Sigma^+ p}_s$   &$3.74$  &$3.60$   &$3.53$   &$3.63$ & $3.31$ & $3.14$\\
$a^{\Sigma^+ p}_t$   &$0.70$  &$0.65$   &$0.60$   &$0.56$ & $0.29$ & $-0.25$\\
$r^{\Sigma^+ p}_t$  &$-2.14$ &$-2.78$ & $-3.55$ & $-4.36$ & $-11.54$ & $-25.35$\\
\hline
$(^3_\Lambda \rm H)$ $E_B$ &$-2.35$ &$-2.34$ &$-2.34$ &$-2.36$  & $-2.27$ & $-2.30$\\
\hline
\end{tabular}
\end{table}
The magnitudes of the $\Lambda p$ singlet and triplet scattering lengths obtained
within chiral EFT are smaller than the corresponding values of the 
J{\"u}lich '04 and Nijmegen NSC97f models (last two columns),
which is also reflected in the small $\Lambda p$ cross 
section near threshold, see Fig. \ref{fig:6.0}a. 
But despite of this significant difference the EFT interaction yields a correctly 
bound hypertriton too, see last row in Table \ref{tab:6.2}. 
The singlet $\Sigma^+ p$ scattering length predicted by chiral EFT is about half as 
large as the values found for the meson-exchange $YN$ potentials. Like in the latter models and 
other $YN$ interactions, the value of the triplet $\Sigma^+ p$ scattering length obtained by
chiral EFT is fairly small. Contrary to NSC97f, but as in the J\"ulich '04 $YN$ model, 
there is repulsion in this partial wave.

\begin{figure}
\resizebox{\textwidth}{!}{%
  \includegraphics*[2.0cm,16.9cm][10.75cm,27cm]{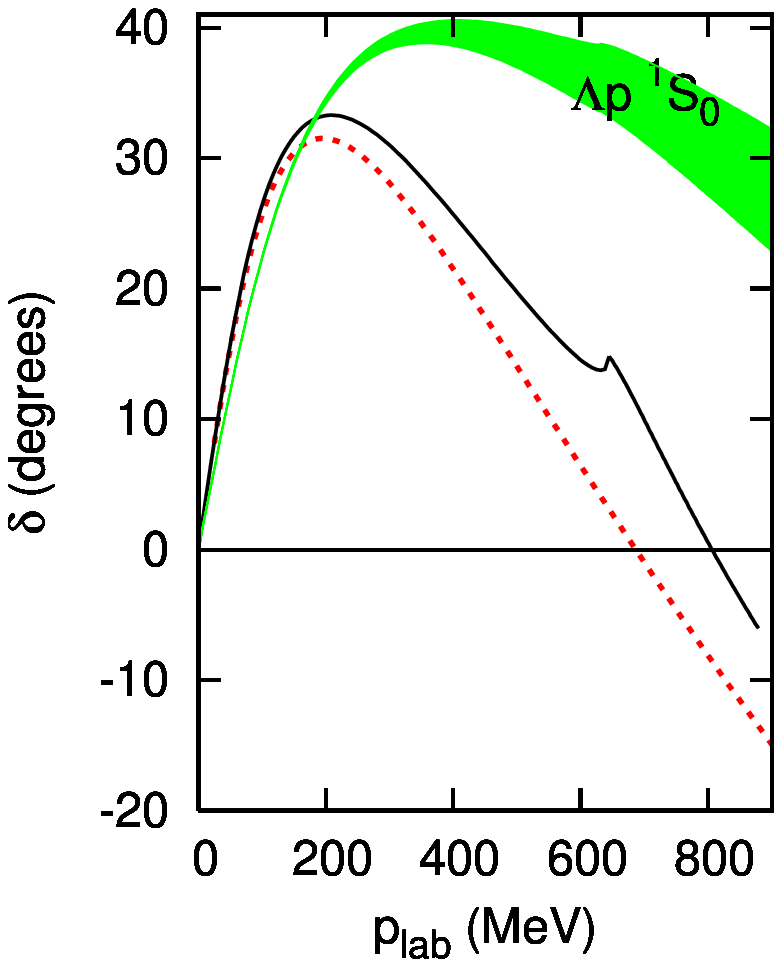}
  \includegraphics*[2.0cm,16.9cm][10.75cm,27cm]{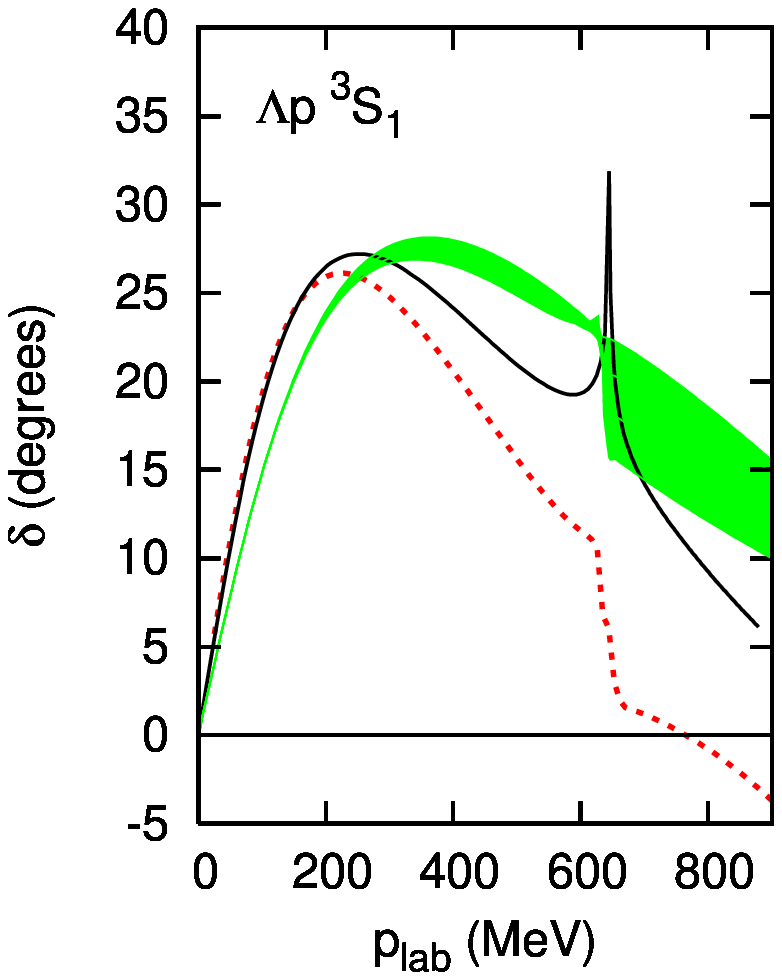}
  \includegraphics*[2.0cm,16.9cm][10.75cm,27cm]{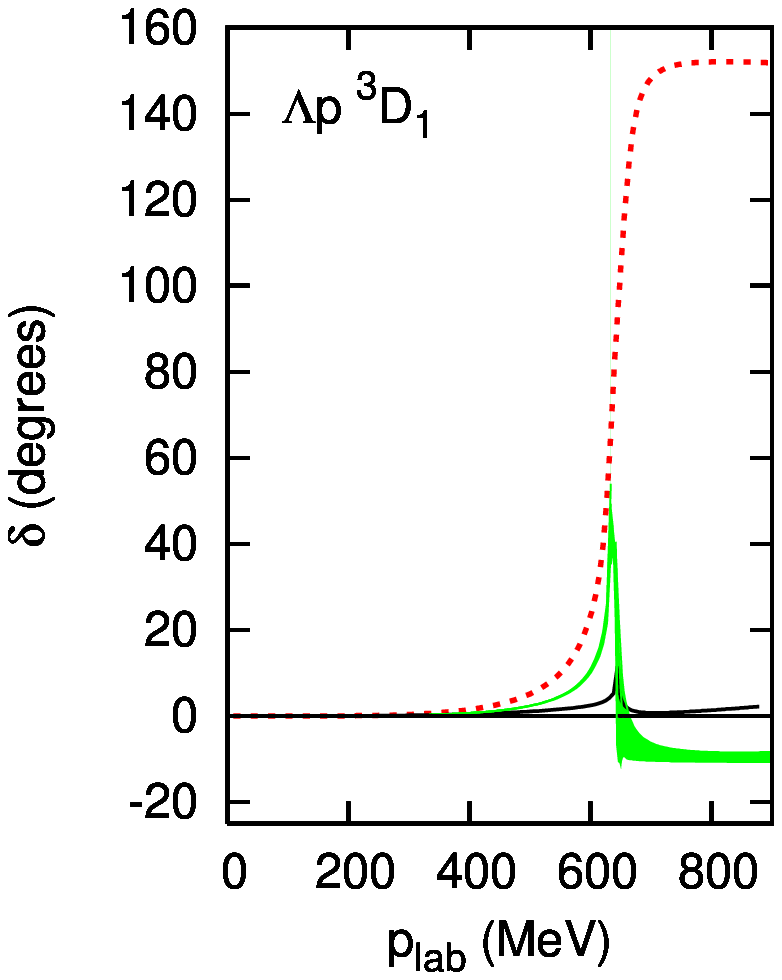}
}
\hfill \break
\resizebox{\textwidth}{!}{%
  \includegraphics*[2.0cm,16.9cm][10.75cm,27cm]{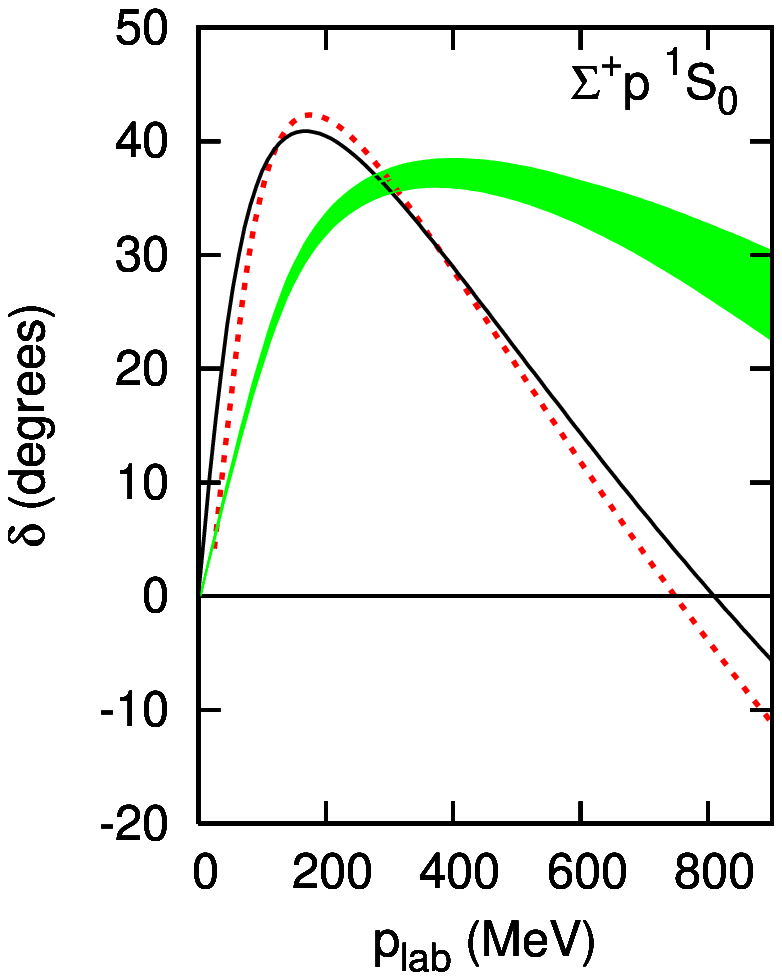}
  \includegraphics*[2.0cm,16.9cm][10.75cm,27cm]{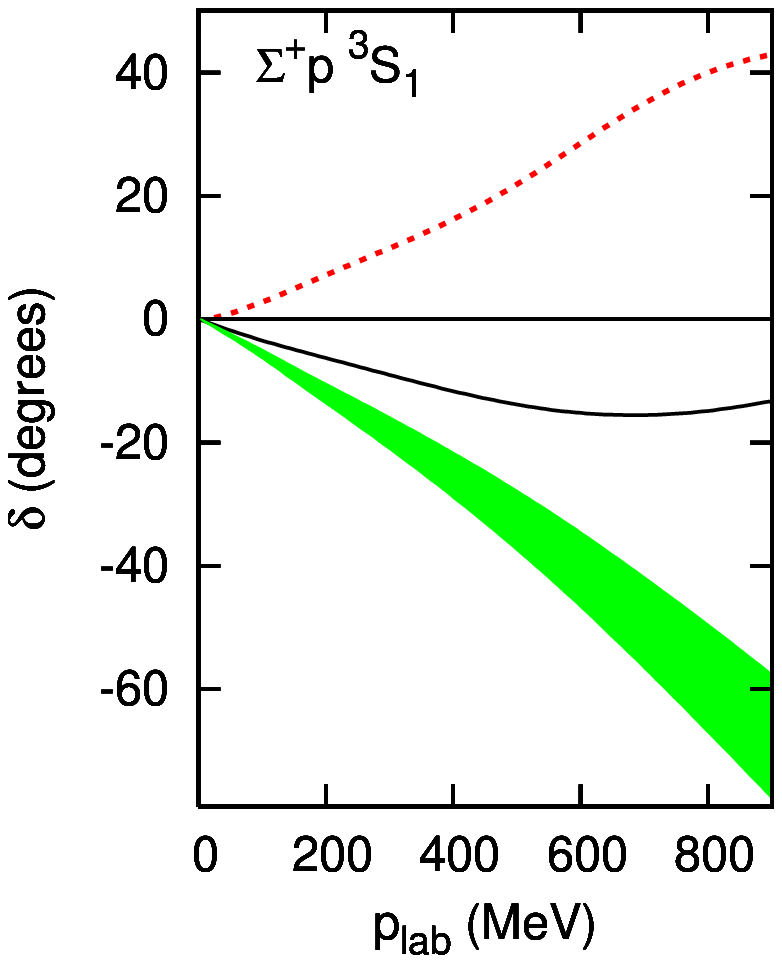}
  \includegraphics*[2.0cm,16.9cm][10.75cm,27cm]{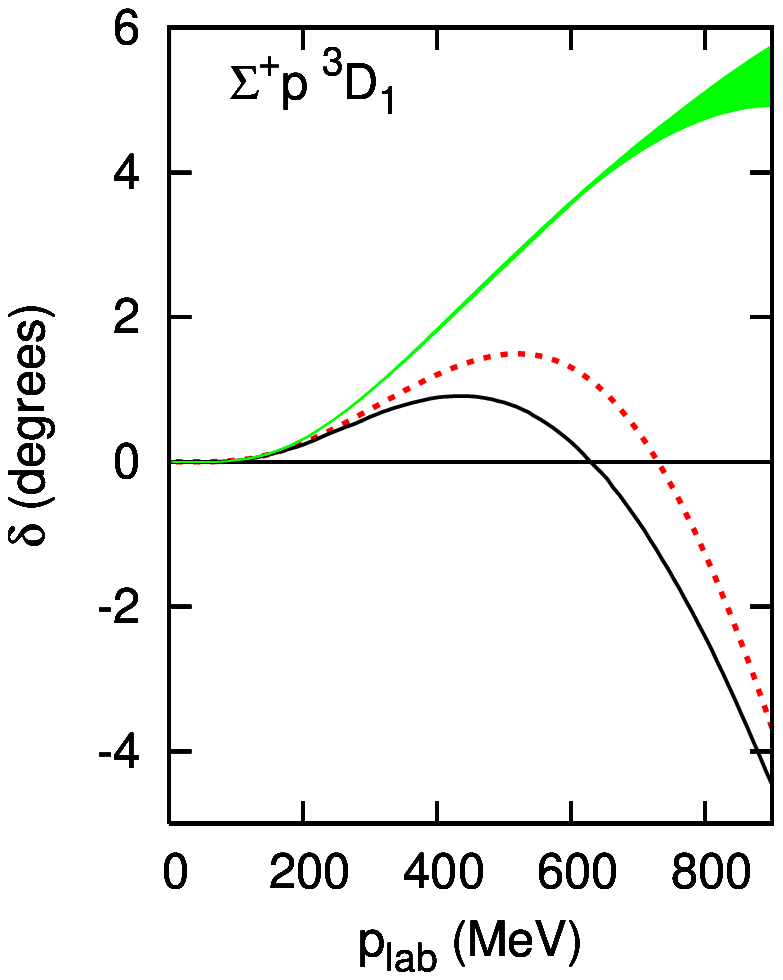}
}
\caption{The $\Lambda p$ and $\Sigma^+ p$ $S$-wave and ${}^3D_1$ phase shifts $\delta$ as a function of $p_{{\rm lab}}$. 
Since the phases of the J{\"ulich '04 model are calculated in the isospin basis, its $\Sigma N$} threshold does not exactly coincide with the others. Same description of curves as in Fig. \ref{fig:6.0}.}
\label{fig:6.4}
\end{figure}

Some $S$- and $D$-wave phase shifts for $\Lambda p$ and $\Sigma^+p$ are shown in Fig.~\ref{fig:6.4}. 
As mentioned before, the 
limited accuracy of the $YN$ scattering data does not allow for a unique phase shift analysis. 
This explains why the chiral EFT phase shifts are quite different from the phase shifts of the 
new meson-exchange $YN$ interaction of the J\"ulich group but also from all models presented in 
Ref.~\cite{Rij99}. Indeed, the predictions of the various meson-exchange models also differ 
between each other in most of the partial waves. 
In both the $\Lambda p$ and $\Sigma^+ p$ ${}^1S_0$ and ${}^3S_1$ partial waves, the LO chiral EFT 
phase shifts are much larger at higher energies than the phases of the meson-exchange models. 
But this is not surprising. First we want to remind the reader that 
the empirical data $YN$ considered in the fitting procedure are at lower energies. 
Second, 
also for the $NN$ interaction in leading order these partial waves were much larger than the 
Nijmegen phase shift analysis, see \cite{Epe00a}. It is expected that this problem for the $YN$ 
interaction can be solved by the derivative contact terms in a NLO calculation, just like in the 
$NN$ case. It is interesting to see that the ${}^3S_1$ $\Sigma^+ p$ phase shift is repulsive in 
chiral EFT as well as in the new J\"ulich meson-exchange model, but attractive in the 
Nijmegen NSC97f model. This has consequences for the $\Sigma^+p$ differential
cross section because, depending on the sign, the interference of the hadronic amplitude
with the Coulomb amplitude differs, cf. Fig.~\ref{fig:6.2}. Unfortunately, the limited 
accuracy of the available $\Sigma^+p$ data does not allow to discriminate between these 
two scenarios. 
 
Results for $P$-wave phase shifts can be found in Refs.~\cite{Rij99,Hai05,Pol06}. Here we just 
want to remark that in case of LO chiral EFT the $P$-waves are the result of pseudoscalar 
meson exchange alone, since we only have contact terms in the $S$-waves in that order. 
Also, contrary to conventional meson-exchange models, in LO chiral EFT there are no spin 
singlet to spin triplet transitions, because of the potential form in Eqs. (\ref{eq11})
and (\ref{eq:3.10}). 
Although the ${}^3D_1$ $\Lambda p$ phase shift near the $\Sigma N$ threshold 
rises quickly for our $YN$ interactions, cf. Fig.~\ref{fig:6.4}, it does not go 
through 90 degrees in both cases -- unlike the Nijmegen model NSC97f \cite{Rij99}. 
The opening of the 
$\Sigma N$ channel is also clearly seen in the ${}^3S_1$ $\Lambda p$ partial wave
for all considered interactions, but again there are significant differences in the 
concrete behavior.

Very recently the chiral EFT model has been employed in Faddeev-type investigations 
of the four-body systems $^4_\Lambda$H and $^4_\Lambda$He \cite{Nog06a}. 
The binding energies of these hypernuclei are especially interesting predictions.
It is has been very difficult in the past to describe their charge symmetry breaking 
(CSB) and the splitting of the $0^+$ ground and $1^+$ 
excited state at the same time \cite{Nogga:2001ef}. In Table~\ref{tab:hypnucl}, we show 
the differences of the binding 
energies of the core nucleus and the hypernucleus,
the $\Lambda$ separation energies, since these are only mildly dependent 
on the $NN$ interaction model used for the calculations \cite{Nogga:2001ef}.
We compare the $\Lambda$ separation energies based on chiral EFT 
and the two considered meson-exchange $YN$ interactions to the 
experimental numbers. It is seen that the separation energies for the excited 
states are somewhat dependent on the cut--off value chosen. Certainly, contributions 
from higher order will be sizable for these observables. However, within 
these uncertainties, the results agree remarkably well with the experimental 
separation energies, which is somewhat less the case for the meson-exchange 
potentials. The CSB of the separation energies is not well described by all 
of the interactions. The Nijmegen model NSC97f includes explicit CSB in the potential,
which induces a sizable but too small effect on the separation energies. 
It will be interesting to study this observable in NLO of the chiral interaction, 
where first explicitly CSB terms contribute.

\begin{table}[tb]
\caption{$\Lambda$ separation energies of the $0^+$ ($E_{sep}(0^+)$) 
      and $1^+$ ($E_{sep}(1^+)$) states and their difference $\Delta E_{sep}$ 
      for $^4_\Lambda$H and the 
      difference of the separation energies for the mirror hypernuclei 
      $^4_\Lambda$He and $^4_\Lambda$H (CSB-$0^+$ and CSB-$1^+$). Results for the 
      chiral EFT $YN$ interaction for various cut--offs $\Lambda$ are compared 
      to predictions for the J\"ulich '04 and Nijmegen NSC97f meson-exchange models
      and the experimental values \cite{Gibson:1995an}.}
\vspace{0.2cm}
\centering
\begin{tabular}{|l|cccc|c|c|c|}
\hline 
& \multicolumn{4}{|c|}{EFT '06} & \, {J\"ulich '04}\, & \, NSC97f \, & \, Expt. \, \\
~$\Lambda$ [MeV] & 500 & 550 & 650 & 700 & & & \cr 
\hline 
~$E_{sep}(0^+)$ [MeV] & 2.63 & 2.46 & 2.36 & 2.38 & 1.87 & 1.60 & 2.04 \cr 
~$E_{sep}(1^+)$ [MeV] & 1.85 & 1.51 & 1.23 & 1.04 & 2.34  & 0.54 & 1.00 \cr 
~$\Delta E_{sep}$ [MeV] & 0.78 & 0.95 & 1.13 & 1.34 & -0.48 & 0.99 & 1.04 \cr 
 \hline 
~CSB-$0^+$ [MeV] & 0.01 & 0.02 & 0.02 & 0.03      & -0.01 & 0.10 & 0.35 \cr 
~CSB-$1^+$ [MeV] & -0.01 & -0.01 & -0.01 & -0.01 & --- &     -0.01    & 0.24 \cr 
\hline 
\end{tabular}
\label{tab:hypnucl}
\end{table}

\section{Summary and outlook}
\label{chap:8}

In this review we presented results based on two different approaches
to the $YN$ interaction, namely on the traditional meson-exchange picture
and on chiral effective field theory. 
 
As far as meson-exchange models of the $YN$ interaction are concerned
we focussed on the recent model of the J\"ulich group, whose 
main new feature is that the contributions both in the
scalar-isoscalar ($\sigma$) and the vector-isovector ($\rho$)
channels are constrained by a microscopic model of correlated
$\pi\pi$ and $K\anti{K}$ exchange.
Besides those contributions from correlated $\pi\pi$ and $K\anti{K}$ exchange
this model incorporates also the standard one-boson exchanges
of the lowest pseudoscalar and vector meson multiplets with coupling
constants fixed by SU(6) symmetry relations. Thus, the long- and intermediate-range 
part of this $YN$ interaction model is completely determined -- either by SU(6) 
constraints or by correlated $\pi\pi$ and $K\anti{K}$ exchange.

The $YN$ interaction derived within chiral EFT is based on a modified Weinberg 
power counting, analogous to the $NN$ force in \cite{Epe05}. The symmetries of 
QCD are explicitly incorporated. Also here it is 
assumed that the interactions in the various $YN$ channels are related via ${\rm
SU(3)}_f$ symmetry. However, since we have done our study in leading order, 
in which the $NN$ interaction can not be described well, we do not connect the present
$YN$ interaction with the $NN$ sector, but focus on the $YN$ system only.

To be specific, the leading-order potential consists of two pieces: firstly, the 
longer-ranged 
one-pseudo{\-}scalar-meson exchanges, related via ${\rm SU(3)}_f$ symmetry in 
the well-known way and secondly, the shorter ranged four-baryon contact term 
without derivatives. The latter contains five independent low-energy constants
that need to be determined from the empirical data. We fixed those five free 
parameters by fitting to 35 low-energy $YN$ scattering data. 
The reaction amplitude is obtained by solving a regularized Lippmann-Schwinger
equation for the chiral EFT interaction. The regularization is done by 
multiplying the strong potential with an exponential regulator function where 
we used a cut--off in the range between $550$ and $700$ MeV. 

The meson-exchange picture has been already applied successfully to the
$YN$ system in the past by many authors. Thus, it is not surprising that a good 
reproduction of the data could be achieved within this approach. 
But it is rather reassuring to see that also chiral effective field theory
works remarkably well for the $YN$ interaction, in particular since we 
have, so far, restricted ourselves to lowest order only.  
Indeed, we could obtain a rather good description of the empirical data, as is
reflected in the total $\chi^2$ which is the range between $28.3$ and $34.6$ 
for a cut--off in the range between $550$ and $700$ MeV. 
In addition 
low-energy differential cross sections and higher energy cross sections, that were not 
included in the fitting procedure, were predicted quite well. 

In a first application to few-baryon systems involving strangeness 
we found that the chiral EFT yields a correctly bound hypertriton \cite{Nog06}. We did not explicitly 
include the hypertriton binding energy in the fitting procedure, but we have fixed the relative 
strength of the $\Lambda N$ singlet and triplet $S$-waves in such a way that a bound hypertriton 
could be obtained. It is interesting to note that a $\Lambda p$ singlet scattering length of $-1.9$ fm 
leads to the correct binding energy. Meson-exchange $YN$ models that yield comparable results 
for the hypertriton binding energy predict here singlet scattering lengths that are typically in
the order of $-2.5$ fm. 

In conclusion, our results strongly suggest that the chiral effective field theory scheme, applied 
in Ref.~\cite{Epe05} to the $NN$ interaction, also works well for the $YN$ interaction. In the future 
it will be interesting to study the convergence 
of the chiral EFT for the $YN$ interaction by doing NLO and NNLO calculations. 
In particular a combined $NN$ and $YN$ study in chiral EFT, starting with a NLO calculation, 
needs to be performed. 
Also an SU(3) extension to the hyperon-hyperon ($YY$) sector is of interest. In this case only
one additional low-energy constant arises within the EFT approach in LO. This constant 
could be fixed by available data on the reaction $\Xi^- p \to \Lambda\Lambda$ \cite{Ahn06}, 
say, and then predictions can be made for all reaction channels in the strangeness -2 sector. 
In particular, one would then be able to obtain an estimate for the $\Lambda\Lambda$ interaction, 
whose strength is rather crucial for the existence of doubly strange hypernuclei. 
With regard to the interactions presented in this review it will be interesting to see 
their performance when employed in further calculations of strange few-baryon systems 
as well as in hypernuclei. For example, preliminary results for the four-body hypernuclei 
${}^4_\Lambda {\rm H}$ and ${}^4_\Lambda {\rm He}$ show that the 
chiral EFT predicts reasonable $\Lambda$ separation energies for ${}^4_\Lambda {\rm H}$, 
though the charge dependence of the $\Lambda$ separation energies is not reproduced
(as expected at lowest order). 

 \bibliographystyle{phaip}
 \bibliography{jhyn}
%


\printindex
\end{document}